\documentclass[a4paper,fleqn,usenatbib]{mnras}

\usepackage{newtxtext,newtxmath}

\usepackage[T1]{fontenc}
\usepackage{ae,aecompl}

\usepackage{graphicx}	
\usepackage{amsmath}	

\usepackage{adjustbox} 
\usepackage{lipsum}
\usepackage{caption} 
\usepackage[para,online,flushleft]{threeparttable} 

\newcommand{\Me}{M$_\oplus$}
\newcommand{\Mjup}{M$_\mathrm{Jup}$}
\newcommand{\Msun}{M$_{\sun}$}
\newcommand{\Lsun}{$L_{\sun}$}

\newcommand{\ap}{a_\mathrm{p}}
\newcommand{\ep}{e_\mathrm{p}}

\newcommand{\Md}{M_\mathrm{d}}
\newcommand{\rmin}{r_{\min}}

\newcommand{\dgap}{\delta_\mathrm{g}}

\newcommand{\wgap}{w_\mathrm{g}}
\newcommand{\rgap}{r_\mathrm{g}}

\newcommand{\rc}{r_\mathrm{c}}

\newcommand{\uJybeam}{$\mu$Jy~beam$^{-1}$}
\newcommand{\kms}{km~s$^{-1}$}

\title[ALMA observations of HD206893]{Insights into the planetary dynamics of HD~206893 with ALMA}
\author[S. Marino et al.]{ S. Marino$^{1}$\thanks{E-mail:
    sebastian.marino.estay@gmail.com}, A. Zurlo$^{2}$, V. Faramaz$^{3}$, J. Milli$^{4}$, Th. Henning$^{1}$, G. M. Kennedy$^{5,6}$, \newauthor{L. Matr\`a$^{7}$, S. P\'erez$^{8}$, P. Delorme$^{9}$, L. A. Cieza$^{2}$ and A. M. Hughes$^{10}$} 
  \\
  $^{1}$Max Planck Institute for Astronomy, K\"onigstuhl 17, 69117 Heidelberg, Germany\\
  $^{2}$N\'ucleo de Astronom\'ia, Facultad de Ingenier\'ia y Ciencias, Universidad Diego Portales, Av. Ejercito 441, Santiago, Chile\\
  $^{3}$Jet Propulsion Laboratory, California Institute of Technology, 4800 Oak Grove drive, Pasadena CA 91109, USA.\\
  $^{4}$Univ. Grenoble Alpes, CNRS, IPAG, 38000 Grenoble, France\\
  $^{5}$Department of Physics, University of Warwick, Gibbet Hill Road, Coventry, CV4 7AL, UK\\
  $^{6}$Centre for Exoplanets and Habitability, University of Warwick, Gibbet Hill Road, Coventry, CV4 7AL, UK\\
  $^{7}$School of Physics, National University of Ireland Galway, University Road, Galway, Ireland\\
  $^{8}$Departamento de F\'isica, Universidad de Santiago de Chile, Av. Ecuador 3493, Estaci\'on Central, Santiago, Chile\\
  $^{9}$Univ. Grenoble Alpes, CNRS, IPAG, 38000 Grenoble, France\\
$^{10}$Astronomy Department and Van Vleck Observatory, Wesleyan University, Middletown, CT 06459, USA}

\date{Accepted XXX. Received YYY; in original form ZZZ}

\pubyear{2020}

\begin{document}
\label{firstpage}
\pagerange{\pageref{firstpage}--\pageref{lastpage}}
\maketitle

\begin{abstract}
  
  Radial substructure in the form of rings and gaps has been shown to
  be ubiquitous among protoplanetary discs. This could be the case in
  exoKuiper belts as well, and evidence for this is emerging. In this
  paper we present ALMA observations of the debris/planetesimal disc
  surrounding HD~206893, a system that also hosts two massive
  companions at 2 and 11~au. Our observations reveal a disc extending
  from 30 to 180~au, split by a 27 au wide gap centred at 74~au, and
  no dust surrounding the reddened brown dwarf (BD) at 11~au. The gap
  width suggests the presence of a 0.9~\Mjup\ planet at 74~au, which
  would be the third companion in this system. Using previous
  astrometry of the BD, combined with our derived disc orientation as
  a prior, we were able to better constrain its orbit finding it is
  likely eccentric ($0.14^{+0.05}_{-0.04}$). For the innermost
  companion, we used RV, proper motion anomaly and stability
  considerations to show its mass and semi-major axis are likely in
  the range 4--100 \Mjup\ and 1.4--4.5~au. These three companions will
  interact on secular timescales and perturb the orbits of
  planetesimals, stirring the disc and potentially truncating it to
  its current extent via secular resonances. Finally, the presence of
  a gap in this system adds to the growing evidence that gaps could be
  common in wide exoKuiper belts. Out of 6 wide debris discs observed
  with ALMA with enough resolution, 4--5 show radial substructure in
  the form of gaps.
  
\end{abstract}





\section{Introduction}
\label{sec:intro}

The study of exoplanetary systems has been revolutionised in the last
decade with the discovery of thousands of exoplanets and several
hundreds of debris discs (analogous to the Kuiper belt), evidenced by
short-lived dust that is being replenished via collisions among an
underlying population of planetesimals \citep[see reviews
  by][]{Wyatt2008, Hughes2018}. Some of these systems are known to
host both exoplanets and \textit{exoKuiper} belts, allowing for a more
detailed characterisation of their architecture, dynamics and
formation since they provide complementary information
\citep[e.g.][]{Moro-Martin2010, Moro-Martin2007}.

As the number of known systems hosting both planets and exoKuiper
belts grew, studies have tried to find correlations between the
two. Some have provided tentative evidence of a possible higher
occurrence rate of debris discs (indicative of more mass in the form
of planetesimals) in systems hosting low-mass planets detected through
radial velocities \citep[RV,][]{Wyatt2012, Marshall2014,
  Moro-Martin2015}, typically located within 1~au and with discs at
tens of au \citep[e.g.][]{Kennedy2015superearths, Marino201761vir},
but this trend has been recently shown to be not significant
\citep{Yelverton2020}. On the other hand, there seems to be an
anticorrelation between the presence of massive close-in planets (or
stellar metallicity) and detectable debris discs \citep{Greaves2004,
  Moro-Martin2007}. More recently, \cite{Meshkat2017} also showed that
systems with bright debris discs seem to be more likely to have
planets at least a few times more massive than Jupiter at separations
of 10--1000~au, where planets and debris generating planetesimals
could interact. The origin for these tentative correlations is still
unclear and it is likely that many factors during the planet formation
process and subsequent dynamical evolution contribute to these.




One way to improve our understanding is to look in detail how planets
and debris discs interact. Thanks to ALMA it has been possible to
image tens of debris discs at millimetre wavelengths, typically
tracing mm-sized grains unaffected by radiation \citep{Burns1979} or
gas drag forces \citep[e.g.][and references therein]{Marino2020gas},
thus tracing the spatial distribution of the parent km-sized
planetesimals. ALMA images have revealed at unprecedented detail
asymmetric structures \citep[e.g. $\beta$~Pic, Fomalhaut and
  HD~202628,][]{Dent2014, MacGregor2017, Faramaz2019}, annular gaps
\citep[HD~107146, HD~92945, HD~15115,][]{Ricci2015, Marino2018hd107, Marino2019,
  MacGregor2019}, and vertical substructure
\citep[e.g. $\beta$~Pic][]{Matra2019betapic}, suggesting the presence
of as-yet unseen low mass planets.

While most systems with exoKuiper belts do not have known planetary
mass companions, in a few of these it has been possible to directly
image one, thus enabling the study of planet-disc interactions in more
detail. There are well known examples such as $\beta$~Pic with a
massive planet possibly warping the disc \citep{Mouillet1997,
  Lagrange2012, Lagrange2019, Matra2019betapic}; HR~8799 with four
giant planets creating a scattered disc and possibly replenishing its
warm dust closer in \citep[e.g.][Faramaz et al. in prep]{Marois2010,
  Zurlo2016, Booth2016, Read2018, Wilner2018, Geiler2019}; HD~95086's
axisymmetric disc implying a low eccentricity of its 4~\Mjup\ planet
\citep{Rameau2016, Su2017}; and Fomalhaut having a narrow and
eccentric planetesimal belt \citep{Kalas2005, Boley2012, Acke2012,
  MacGregor2017}, implying that its candidate companion on an
eccentric orbit has a low mass ($\sim$Earth or super-Earth) and is not
sculpting the belt \citep{Quillen2006, Kalas2008, Chiang2009,
  Beust2014, Faramaz2015}, or is not a compact object but rather the
dusty aftermath of a recent planetesimal collision
\citep{Gaspar2020}. Some exoKuiper belt host systems even have
companions in the brown dwarf or low stellar mass regime, suggesting
that their likely formation through gravitational instability
\citep[][]{Boss1997, Boss2003, Boss2011, Vorobyov2013} is compatible
with the formation of massive Kuiper belt analogues, e.g. HR~2562
\citep{Konopacky2016}, HD~193571 \citep{MussoBarcucci2019} HD~92536
\citep{Launhardt2020}, and HD~206893 \citep{Milli2017hd206}. The
latter is the subject of this paper. For even more massive companions,
\cite{Yelverton2019} found a significant lower detection rate of
debris discs around binaries, with no discs detected in binaries with
separations between 25--135~au \citep[comparable to typical debris
  disc radii,][]{Matra2018mmlaw}. This is likely due to dynamical
perturbation inhibiting planetesimal formation or clearing any debris
disc formed near those separations.


Located at 40.8~pc \citep{Gaiadr2}, the F5V star HD~206893 is known to
host a companion, HD~206893~B, at a separation of $\sim11$~au
\citep{Milli2017hd206} and a debris disc \citep{Moor2006, Chen2014}
that was marginally resolved with \textit{Herschel}
\citep{Milli2017hd206}. Given the estimated age of this system of
50--700 Myr, the companion mass is probably in the range
12--50~\Mjup\ \citep{Delorme2017}, and thus it is likely a brown dwarf
(BD). Astrometric follow up of this companion using VLT/NACO and
SPHERE placed some constraints on the period (or semi-major axis),
orientation of the orbit, and set an upper limit to the eccentricity
of $\sim0.5$ \citep{Grandjean2019}. The same study using
\textit{Hipparcos} \citep{vanLeeuwen2007} and \textit{Gaia} DR2 data
\citep{Gaiadr2} revealed a significant proper motion anomaly in a
direction which cannot be explained by the BD, suggesting the presence
of an additional companion closer-in \citep[also confirmed
  in][]{Kervella2019}. This additional companion would also be
responsible for an observed RV drift (or stellar acceleration),
indicating that this inner companion (HD~206893 C) must have at least
a mass of $\sim15$~\Mjup\ and orbit with a semi-major axis between
1.4~au to 2.6~au \citep{Grandjean2019}. These two companions make this
system an ideal target to image its disc with ALMA, in order to both
constrain the dynamics of this system and look for additional
companions that could shape the distribution of planetesimals. This
paper is structured as follows. In \S\ref{sec:obs} we present our new
ALMA observations that show evidence of a gap. We then fit these
observations using a parametric disc model in \S\ref{sec:parmodel} to
constrain the disc orientation and radial structure. Using these
constraints, particularly the disc orientation, in \S\ref{sec:orbit}
we improve the previous orbital constraints of HD~206893~B by assuming
it is co-planar with the disc. In \S\ref{sec:dis} we discuss and
summarise the different constraints on companions in this system and
potential origins of the gap. Finally, in \S\ref{sec:conclusions} we
summarise our results and conclusions.

\section{Observations}
\label{sec:obs}

We observed HD~206893 with ALMA in band 7 (average wavelength and
frequency of 0.88 mm and 342~GHz) as part of the cycle 5 project
2017.1.00828.S (PI: A. Zurlo). Observations were taken both with the
Atacama Compact Array (ACA) and the main 12m array, in order to
recover the large scale structure up to sizes of $20\arcsec$ and the
small scale structures down to $0\farcs3$, respectively. Details about
these observations are summarised in Table \ref{tab:obs}. The
correlator was set up using four spectral windows to study primarily
the dust continuum emission in the system. Three of these were centred
at 348.4, 336.5 and 334.6~GHz, with a bandwidth of 2 GHz and a channel
width of 15.6~MHz. The fourth spectral window was centred at 348.4~GHz
and had a bandwidth of 1.875 GHz, with a narrower channel width of
0.488 MHz (0.42~\kms and effective bandwidth of
1.1~\kms\footnote{https://help.almascience.org/index.php?/Knowledgebase/Article/View/29})
to search for a serendipitous CO J=3-2 detection. The data was
calibrated using \textsc{casa}~5.4 \citep{casa} and the standard
calibration routines provided by ALMA. Additionally, we applied a
phase center shift to the ACA data which was not correctly centred on
HD~206893 at the corresponding epoch according to its Gaia DR2
astrometry. To complement these band 7 observations (both continuum
and CO emission), we also retrieved archival band 6 observations
(1.3~mm, 222~GHz) which we calibrated using the standard ALMA routines
for CASA. These observations are described in Nederlander et al
(2020). Below we present the analysis of continuum and CO observations
 
\begin{table*}
  \centering
  \caption{Summary of band 7 (12m and ACA) and band 6 (12m)
    observations. The image rms and beam size correspond to Briggs
    weighting with a robust parameter of 2.0.}
  \label{tab:obs}
  \begin{adjustbox}{max width=1.0\textwidth}
    \begin{tabular}{lrcccc} 
  \hline
  \hline
  Observation & Dates              & t$_\mathrm{sci}$  & Image rms  & beam size (PA) & Min and max baselines  \\
                &                  &      [hours]       & [$\mu$Jy]  &                &  [m] (5th and 95th percentiles)   \\
  \hline
  Band 7 - 12m & 26, 28, 30 Sep 2018, 20 Apr 2019    & 4.3  & 9.1 & $0\farcs31\times0\farcs25$ ($-82\degr$) & 55 and 733  \\
  Band 7 - ACA & 23 Oct 2017   & 7.9 &  110 &  $5\farcs0\times2\farcs7$ ($88\degr$) & 9 and 45  \\ 
               & 10 Mar, 5, 7 Apr, 8, 14-18, 26 May 2018  & & &&\\
  Band 6 - 12m & 27 Jun, 30 Aug, 10, 17 Sep 2018 & 4.3 & 4.9 &  $0\farcs70\times0\farcs57$ ($67\degr$) & 45 and 670 \\
    \hline
  \end{tabular}
  
  \end{adjustbox}
\end{table*}

 \begin{figure*}
  \centering \includegraphics[trim=0.0cm 0.0cm 0.0cm 0.0cm, clip=true,
    width=0.45\textwidth]{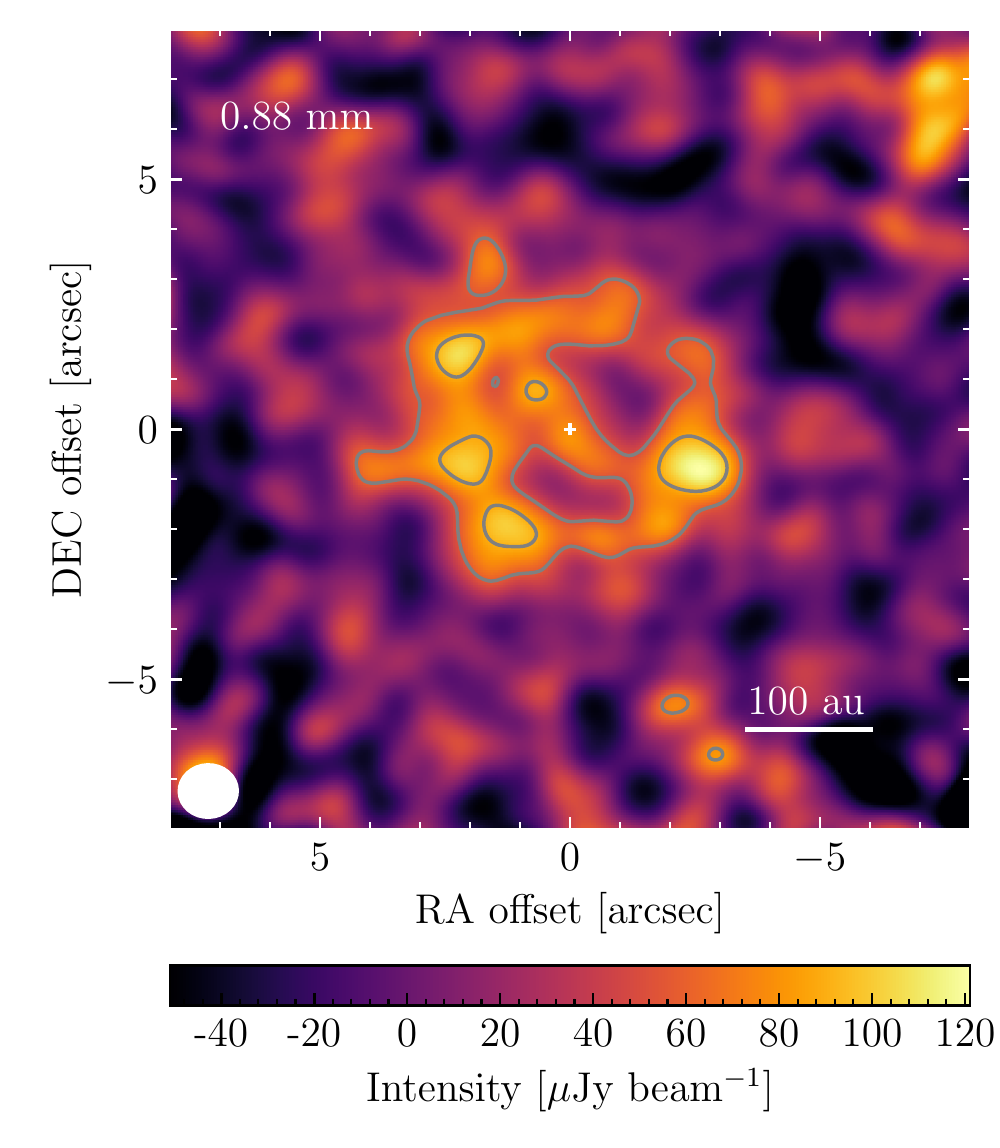}
  \includegraphics[trim=0.0cm 0.0cm 0.0cm 0.0cm, clip=true,
    width=0.45\textwidth]{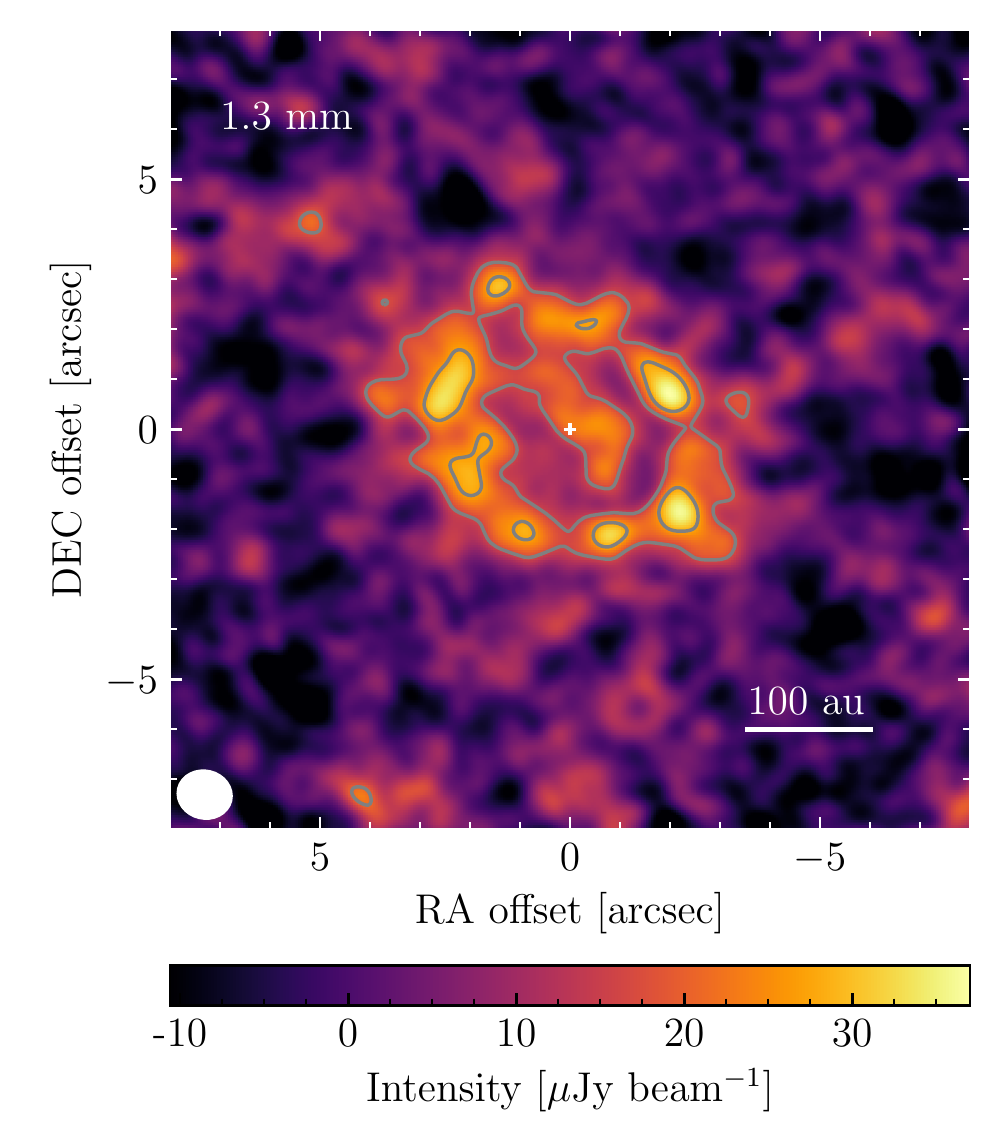}
 \caption{Continuum Clean images at 0.88~mm (12m+ACA, left panel) and
   1.3~mm (right panel) of HD~206893 obtained using Briggs weighting
   and a robust parameter of 2. Additionally, we applied a uv tapering
   of $0\farcs9$ to the band 7 data and $0\farcs4$ to the band 6
   data. The images are also corrected by the primary beam, hence the
   noise increases towards the edges. The contours represent 3 and 5
   times the image rms (17 and 5.3~\uJybeam\ at the center of the band
   7 and 6 images, respectively). The stellar position is marked with
   a white cross near the center of the image (based on Gaia DR2) and
   the beams are represented by white ellipses in the bottom left
   corners ($1\farcs02\times0\farcs91$ and $0\farcs92\times0\farcs80$,
   respectively). }
 \label{fig:alma}
\end{figure*}

\subsection{Dust continuum}
\label{sec:continuum}

Continuum images at wavelengths of 0.88 and 1.3 mm are obtained using
the \textsc{tclean} task in CASA with Briggs weighting and a robust
parameter of 2.0 (to maximise sensitivity). These clean images are
presented in Figure \ref{fig:alma}. The images are smoothed using a
Gaussian tapering\footnote{This is done in the uv space through the
  \textsc{tclean} argument uvtaper.} of $0\farcs9$ at 0.88 mm and
$0\farcs4$ at 1.3 mm, which leads to a loss of resolution, but allows
for an increase of the signal-to-noise per beam. The beam size in the
tapered images is $1\farcs02\times0\farcs91$ at 0.88~mm and
$0\farcs92\times0\farcs80$ at 1.3~mm. Disc emission is detected at
both wavelengths within $4\arcsec$ (160~au) of the star, distributed
over a wide range of radii. At the center of the images the star is
significantly detected (more clearly seen in non-tapered images, see
\S\ref{sec:bddust}). The stellar flux is consistent with
Rayleigh-Jeans extrapolations of its flux at shorter wavelengths
(i.e. 30~$\mu$Jy at 0.88~mm and 13~$\mu$Jy at 1.3~mm), and thus we
attribute it to photospheric emission. We estimate a total integrated
flux of $2.68\pm0.36$ and $1.05\pm0.12$ mJy at 0.88 and 1.3 mm
(including 10\% absolute calibration uncertainties). These fluxes are
computed by integrating all emission within an ellipse of semi-major
axis $5\arcsec$ ($\sim200$~au) and oriented as the disc on the sky
(PA$=61\degr$ and $i=40\degr$, see \S\ref{sec:results}). In both band
6 and band 7 maps, there is evidence of extended emission arising near
the star suggesting that the planetesimal disc is wide. The detail
radial structure is analysed below in \S\ref{sec:radial}.


\subsubsection{Dust radial structure}
\label{sec:radial}
\begin{figure}
  \centering \includegraphics[trim=0.0cm 0.0cm 0.0cm 0.0cm, clip=true,
    width=1.0\columnwidth]{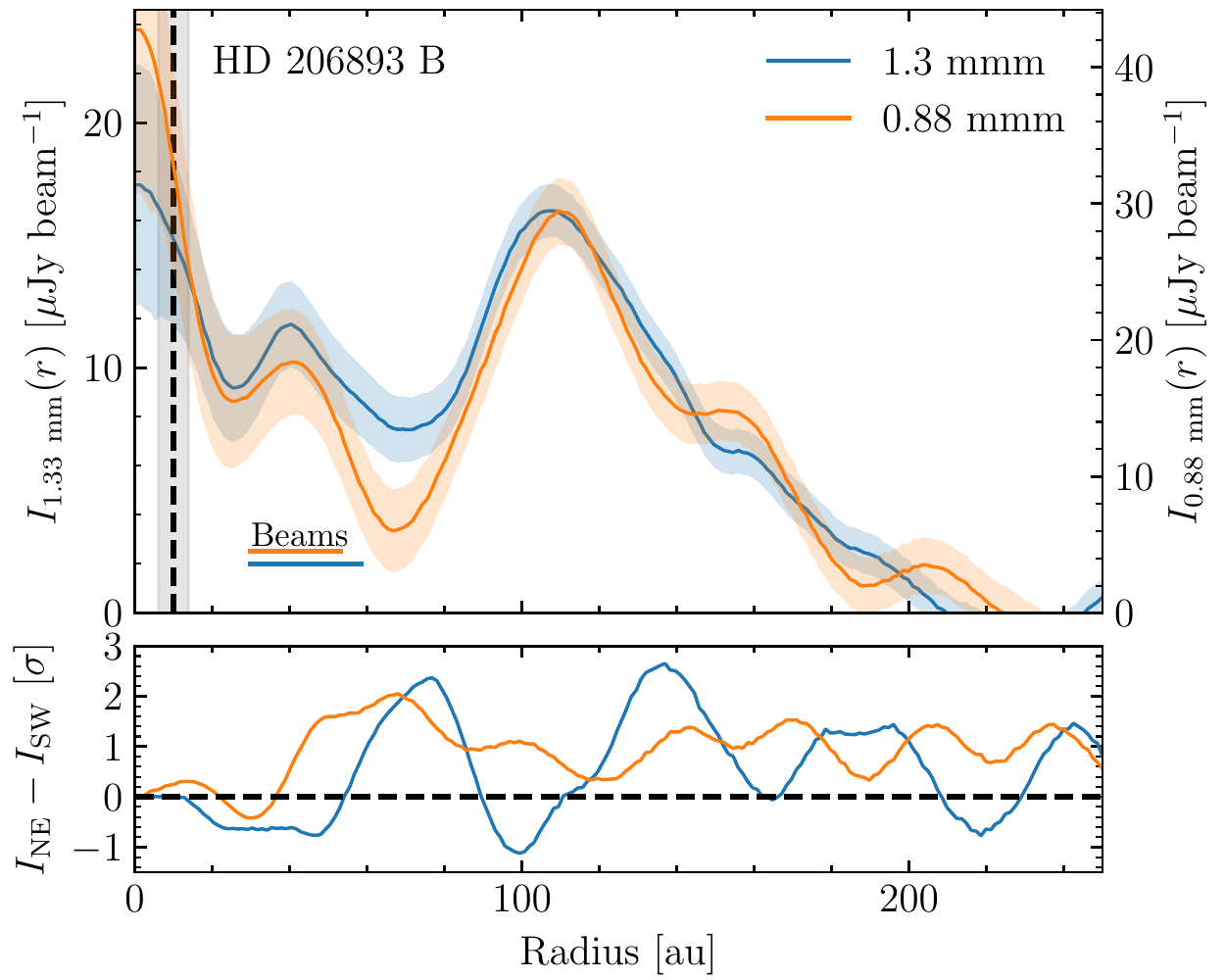}
 \caption{Azimuthally averaged surface brightness profile computed by
   deprojecting the emission according to our best fit model
   (\S\ref{sec:results}) and azimuthally averaging the emission
   (Top). The shaded regions correspond to $1\sigma$
   uncertainties. Note that the shaded regions are representative of
   the uncertainty over a resolution element, i.e. 23~au and 29 for
   band 7 and band 6, respectively. The vertical dashed line
   represents the semi-major axis of B and the grey region its chaotic
   zone if on a circular orbit. The bottom panel shows the difference
   between the North East and South West halves of the disc in
   significance levels (i.e. the difference is divided by the local
   uncertainty).}
 \label{fig:radial}
\end{figure}

In order to study the radial structure of the disc, we azimuthally
average the deprojected emission \citep[as in][]{Marino2016} using the
best fit disc position angle and inclination (see
\S\ref{sec:parmodel}). Both band 6 and band 7 profiles are shown in
Figure \ref{fig:radial}. For this process, we use the band 6 clean
image without tapering and the band 7 image with a $0\farcs4$
tapering. This choice results in a similar beam at both wavelengths
($0\farcs57\times0\farcs50$ at 0.88~mm and $0\farcs70\times0\farcs57$
at 1.3~mm) and is a good compromise between spatial resolution and
signal to noise (see images in \S\ref{appendix}). Based on these
profiles, the disc emission is detected from 30 to 180~au, with a peak
near 110~au and a local minimum at roughly 70~au. This minimum hints
at the presence of a gap at a similar radial distance compared to
HD~107146, HD~92945 and HD15115 \citep[$72\pm3$, $73\pm3$ and
  $59\pm5$~au, respectively,][]{Marino2018hd107, Marino2019,
  MacGregor2019}. The gap seems to be deeper at 0.88~mm, but this is
likely due to the lower resolution at 1.3~mm which does not resolve
well this minimum. Interior to the gap, the disc intensity peaks at
around 40~au, but its exact inner edge is uncertain. Note that the
emission interior to 20 au is simply consistent with photospheric
emission from the star convolved with the beam. Moreover, the disc is
not expected to extend interior to 15~au since it would be truncated
by HD~206893~B's chaotic zone if it is on a circular orbit and has a
mass of 12~\Mjup. Where exactly it should be truncated is uncertain
since B could be more massive (up to 50~\Mjup) and/or on an eccentric
orbit ($<0.5$). Moreover, secular interactions between the two inner
companions could have depleted the disc at regions between 20-40~au
via secular resonances (see discussion in \S\ref{sec:secularres}).

Since some debris discs are known to be asymmetric
\citep[e.g.][]{Dent2014, MacGregor2017, Faramaz2019, Marino2019}, and
expected to be so when interacting with planets on eccentric orbits
\citep[e.g.][]{Pearce2014, Regaly2018}, we perform two tests in order
to search for asymmetries. We first compare the integrated flux of one
half of the disc against the other while varying the angle of the axis
that divides the two halves. We find that when comparing the North
East and South West halves (divided by the disc minor axis), this
difference is maximised and marginally significant ($2\sigma$) at both
wavelengths. Averaging both wavelengths, we find that the North East
side is $30\pm13\%$ brighter. This difference suggests the disc could
be either asymmetric or instead the measured flux is contaminated by a
background sub-millimetre galaxy (SMG) as found in other ALMA
observations of debris discs \citep[e.g.][]{Marino201761vir,
  Zapata2018}. In order to check the radial location where this
difference is strongest, we compute azimuthally averaged radial
profiles of the two disc halves and subtract them (bottom panel of
Figure~\ref{fig:radial}). We find that the NE side is overall brighter
at almost all radii, especially inside the gap and at 140 au, although
these differences are not larger than $3\sigma$. They do reveal
nevertheless that the flux difference does not arise from a single
compact SMG \citep[typically smaller than $0.5\arcsec$ or 20~au at
  HD~206893's distance,][]{Simpson2015size, Lindroos2016,
  Fujimoto2017}, but rather from a much broader region. In
\S\ref{sec:gap} we will discuss what could be the origin of these
asymmetries and how they could be connected with the formation of the
gap and the orbit of HD~206893~B.

\begin{figure}
  \centering \includegraphics[trim=0.0cm 0.0cm 0.0cm 0.0cm, clip=true,
    width=1.0\columnwidth]{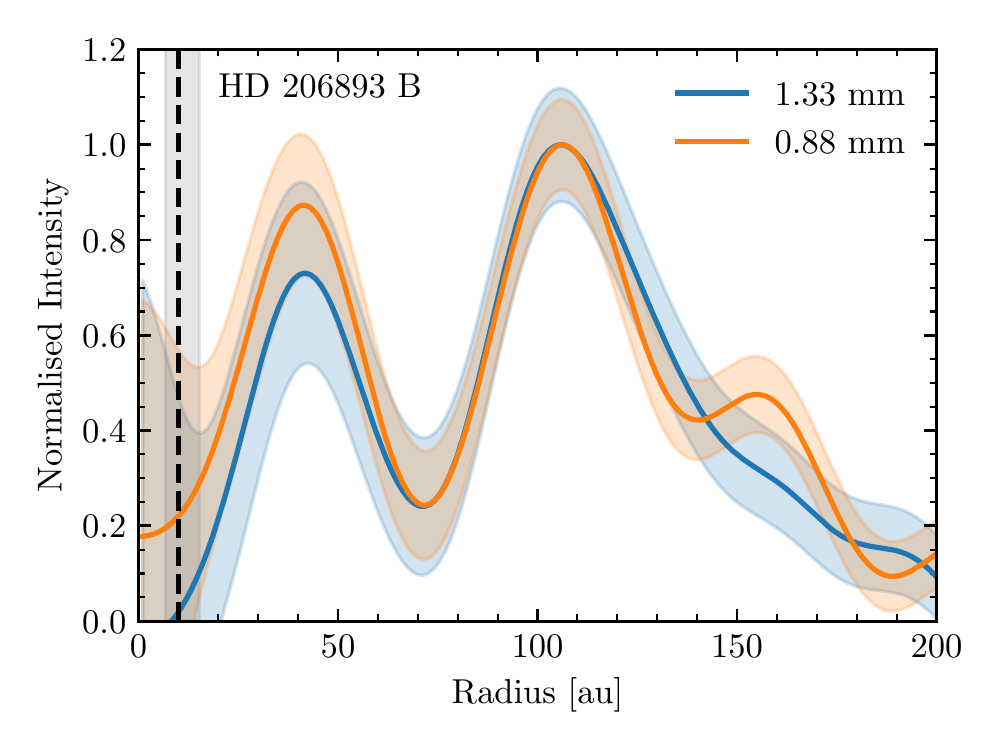}
 \caption{Disc surface brightness profile computed using
   \textsc{Frankenstein} and deprojecting the visibilities according
   to our best fit model (\S\ref{sec:results}). The shaded regions
   correspond to $1\sigma$ uncertainties. The vertical dashed line
   represents the semi-major axis of B and the grey region its chaotic
   zone if on a circular orbit.}
 \label{fig:frank}
\end{figure}

In order to further constrain the radial structure, we use the
\textsc{python} module Frankenstein
\citep[\textsc{frank},][]{Jennings2020} that uses Gaussian process to
reconstruct the intensity radial profile of a disc. We first subtract
the stellar emission from the visibilities by simply subtracting a
constant value of 30~$\mu$Jy at 0.88~mm and 13~$\mu$Jy at
1.3~mm\footnote{the Fourier transform of a point source at the phase
  center is a positive and real constant with a value equal to its
  total flux}. Then, using the derived disc orientation from
\S\ref{sec:results}, we deproject the visibilities and use
\textsc{frank} to reconstruct the radial profile at 0.88 and
1.3~mm. Figure~\ref{fig:frank} presents the derived profiles using a
maximum radius of 400~au and hyperparameters $\alpha=1.01$ and
$w_\mathrm{smooth}=10^{-2}$. These profiles confirm the presence of a
gap centred around 70~au, which appears slightly deeper than in the
profiles derived in the image space due to the higher resolving power
of Frankenstein. These profiles also show a clearer inner cavity of
$\sim20$~au mainly because of the subtraction of the stellar
emission. A caveat in the derivation of these profiles is that no
primary beam correction is taken into account, although its effect is
smaller than 10\% within 150~au in both band 7 and 6 data.


\subsubsection{BD dust emission upper limit}
\label{sec:bddust}

 An additional goal of these observations was to constrain the amount
 of dust surrounding HD~206893~B. \cite{Delorme2017} presented new
 photometric and spectroscopic measurements with SPHERE, and pointed
 out that the brown dwarf companion is a peculiar object. Neither
 empirical models combined with absorption by forsterite nor synthetic
 dusty spectra can describe its very red colour. Since the debris disc
 is outside the orbit of the companion and it has a very low optical
 depth, it cannot explain the reddening and extinction of the BD. A
 possible explanation is that the red colour of this object is
 produced by the extinction from a circumplanetary disc. At NIR
 wavelengths this small disc is obviously unresolved since diffraction
 limited observations lead to 2.0~au resolution at 40~pc in the case
 of SPHERE/IRDIS (K band). However, it could be detected in thermal
 emission at longer wavelength.

\begin{figure}
  \centering \includegraphics[trim=0.0cm 0.0cm 0.0cm 0.0cm, clip=true,
    width=1.0\columnwidth]{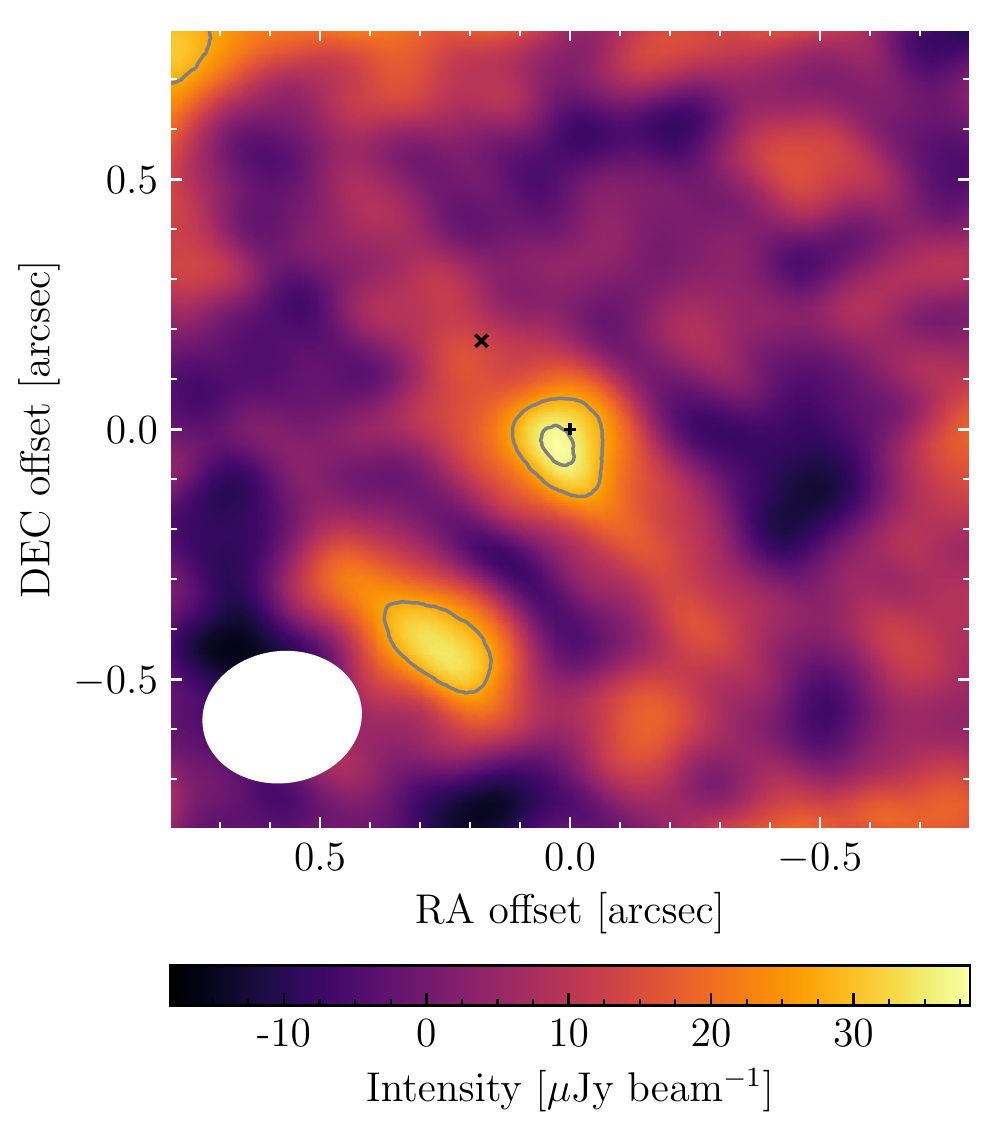}
 \caption{Zoomed clean image at 0.88~mm using Briggs weighting and a
   robust parameter of 2. The contours represent 3 and 4 times the rms
   (9~\uJybeam). The stellar position is marked with a black cross
   near the center of the image (based on Gaia DR2) and the brown
   dwarf position is marked with a black ``x''. The beam is
   represented by a white ellipse in the bottom left corner
   ($0\farcs30\times0\farcs24$).}
 \label{fig:BD}
\end{figure}

 We do not detect any emission arising from a point source towards the
 NE at the BD separation of $\sim0\farcs25$ (see Figure~\ref{fig:BD}),
 where the BD is expected to be\citep{Grandjean2019}. Nevertheless, we
 can place $3\sigma$ upper limits at 0.88 and 1.3 mm of 27 and
 15~$\mu$Jy, respectively. Using the same radiative transfer tools as
 in \cite{Perez2019protolunar}, we estimate a dust mass upper limit of
 $2\times10^{-4}$~\Me\ ($2\times10^{-2}$~$M_\mathrm{moon}$), which is
 equivalent to a dust to planet ratio of $5\times10^{-8}$ for a disc
 with a size $\gtrsim0.1$~au. A much smaller disc would be optically
 thick and could hide a higher mass remaining undetected by
 ALMA. However, this is an unlikely scenario because such a dense and
 compact dusty disc would quickly become depleted in dust due to
 collisional evolution and loss processes (e.g. radiation pressure).

 This non-detection of dust surrounding HD~206893~B does not rule-out
 that B could have a massive circumstellar disc in the form of
 satellites, but that is very depleted in dust. Even if those
 satellites were capable of producing dust through collisions, for
 HD~206893~B that dust would hardly be detectable according to
 theoretical predictions for the collisional evolution of satellite
 swarms \citep{Kennedy2011}. In such swarms, dust is continually
 produced via collisions of km-sized satellites (analogous to
 circumstellar debris discs) and lost due to radiation pressure or
 Poynting-Robertson drag, in which case it is accreted on the
 companion. Given HD~206893's age ($>10$~Myr) and distance (40.8 pc),
 the models by \cite{Kennedy2011} predict fluxes below 1~$\mu$Jy at
 1~mm (see their figure 4 for the case of a system at 10 pc), and thus
 we cannot rule out that HD~206893~B hosts a massive satellite swarm.

 Another possibility to explain the reddening is that it is caused by
 accretion of dust from the circumstellar debris disc. While the
 planetesimal belt resides beyond the orbit of the BD, it is still
 possible that dust is being accreted by the BD, potentially
 explaining its reddened spectrum. Small dust grains are expected to
 migrate in through Poynting-Robertson drag interior to debris discs
 \citep[e.g.][]{Burns1979, Wyatt2005, vanLieshout2014,
   Kennedy2015lbti}, and interact with intermediate planets. Such
 interactions can lead to trapping in resonance, ejections and
 accretion onto the planets \citep{Shannon2015, Bonsor2018}. Using
 equations 21 and 22 in \cite{Bonsor2018}\footnote{There is a typo in
   Table 1 in \cite{Bonsor2018} where $K_\mathrm{ej}$ should be
   $5.14\times10^{8}$ (private communication with Amy Bonsor).} we can
 estimate the rate at which the BD could be accreting dust. We find
 that the BD could be accreting $\mu$m-sized dust at a rate of
 $\sim4$~kg~s$^{-1}$. A simple calculation can show that this level of
 dust accretion cannot lead to the needed reddening, by comparing the
 total cross sectional area accreted over the age of this system vs
 the surface area of the BD. Assuming the accreted dust is all made of
 grains of 1 $\mu$m in radius, have a density of 3~g~cm$^{-3}$, and
 have been accreted for 700 Myr, we find the equivalent dust layer
 would only cover 0.001\% of the surface of the BD, and thus
 insufficient to to produce reddening or extinction at the necessary
 levels \citep[$A_K\sim0.5$,][]{Delorme2017}. Note that to produce
 reddening at NIR larger grains might be necessary and hence the total
 cross-section smaller. Moreover, we neglect here any mixing in the
 atmosphere of the BD that would tend to remove even more of this dust
 below the photosphere, by settling. Therefore we conclude that if
 dust in the atmosphere is indeed responsible for its reddening, it
 cannot be supplied in sufficient quantity by the circumstellar disc.


These findings suggest that the physical conditions within the
atmosphere of HD~206893~B allow to lift dust (or to prevent it to
condense) inside and above its photosphere in quantities that are at
least in the highest ranges that were considered as possible by
substellar atmosphere models. Indeed, \cite{Delorme2017} showed that
the spectra of HD~206893~B could not be correctly fitted without
external extinction, unless the dust settling parameters of the models
were manually tuned to increase dust opacity within the atmosphere
beyond the range that fits other L-dwarfs.



\subsection{CO gas emission}
\label{sec:gas}

We search for any CO line emission in both band 7 (J=3-2) and band 6
(J=2-1) data. This search is done by first subtracting the continuum
emission from the visibilities and imaging the data with
\textsc{tclean}. This produces data cubes with channel widths of
0.43~\kms (effective bandwidth of 1.1~\kms due to Hanning smoothing)
in band 7 and 1.3~\kms (effective resolution of 2.0~\kms due to
Hanning smoothing and averaging every 2 channels) in band 6. The band
7 and 6 cubes have a noise level of 0.7 and 0.3 mJy~beam$^{-1}$ per
channel, respectively, and do not have any significant emission over a
single beam in a single channel. In order to search for low level
emission spread over multiple beams and channels, we applied a
Keplerian mask as in \cite{Matra2015, Marino2016,
  Matra2017fomalhaut}. This procedure corrects for the Doppler shift
in each individual pixel due to Keplerian motion and centers the
emission within a few channels, assuming a disc orientation (derived
in \S\ref{sec:parmodel}) and sense of rotation--- we test both
possible directions. After integrating the emission within a
deprojected radius of 150 au, we still do not detect any significant
emission in the recovered spectra with rms levels of 20 mJy for J=3-2
and 2 mJy for J=2-1. Given this noise level and the fact that real CO
emission in Keplerian rotation would appear as a single peak with a
width equal to the effective bandwidth, we estimate 3$\sigma$ upper
limits for the line fluxes of 66 mJy~\kms for J=3-2 and 12~mJy~\kms
for J=2-1.

These CO flux upper limits can be translated to CO masses assuming
optically thin emission. We cannot simply assume Local Thermodynamic
Equilibrium (LTE) since at these low densities the collisional
excitation of rotational levels can be very low and thus non-LTE
effects must be taken into account \citep{Matra2015}. Instead, we use
the tool developed by \cite{Matra2018} to derive CO mass upper limits
for a wide range of collisional partner densities (spanning from the
radiation dominated regime to LTE) and a range of kinetic temperatues
from 20 to 200~K. We find that our upper limit on CO J=2-1 emission is
the most constraining, with an upper limit of $2.4\times10^{-6}$~\Me.

Since CO gas is expected to be released in collisions if planetesimals
are rich in CO \citep[e.g.][]{Zuckerman2012, Dent2014, Marino2016,
  Matra2017fomalhaut, Moor2017, Kral2017CO, Kral2019, Marino2020gas},
we can use this CO upper limit to place an upper limit in the
CO+CO$_2$ ice mass fraction of planetesimals. In steady-state, CO gas
molecules are photodissociated by interstellar UV photons at a rate
equal to the rate at which they are released from planetesimals. The
latter is expected to be roughly equal to the product of the mass loss
rate of small dust and the mass fraction of CO+CO$_2$ in solids (as
long as CO and CO$_2$ molecules escape before solids are ground down
to $\mu$m-sized grains). Using equation 2 in \cite{Matra2017fomalhaut}
together with the fractional luminosity of the system
($\sim3\times10^{-4}$), stellar mass (1.3~\Msun), stellar luminosity
(2.9~\Lsun), disc radius and width (approximately 70 and 100 au), and
an expected photodissociation rate of 120~yr \citep{Visser2009}, we
find an upper limit of 9\% for the fractional mass of CO+CO$_2$ in
planetesimals. This limit is lower than in some CO-rich comets, but it
is still consistent with the wide distribution of abundances of Solar
System comets \citep{Mumma2011}. Therefore the CO non-detection is
consistent with HD~206893's exoKuiper belt having a volatile
composition similar to that of comets.




\section{Parametric axisymmetric disc model}
\label{sec:parmodel}

In order to quantify the location and width of the gap, we fit a
parametric disc model with a gap as in \cite{Marino2018hd107,
  Marino2019}, combining radiative transfer simulations
\citep[RADMC-3D,][]{radmc3d} and an MCMC fitting procedure in the
visibility space. The surface density of dust is defined by a two
power law distribution, with an inner edge at $\rmin$, surface density
\textit{slopes} (power law index) $\gamma_1$ for $r<r_\mathrm{c}$ and
$\gamma_2$ for $r>r_\mathrm{c}$, and a gap centred at $\rgap$ (with
$\rmin<\rgap<r_\mathrm{c}$) and with a Gaussian profile characterised
by a FWHM $\wgap$. In contrast to \cite{Marino2018hd107, Marino2019},
here we fixed the gap depth to 1 (i.e. the surface density is zero at
$\rgap$). The surface density is normalised such that the total dust
mass is $\Md$.  Note that this dust mass only includes the mass in
grains smaller than 1~cm, which is set by the assumed dust opacities
\citep[see][]{Marino2019}. Since the disc is seen close to face-on and
the total S/N is not very high, there is negligible information about
the disc vertical structure in this data. We thus choose to fix the
vertical extent of the disc to 5\% of the radius and use a uniform
vertical mass distribution to simplify the model and speed up our
simulations. The dust mass together with the assumed opacities set the
disc overall brightness at 0.88~mm, while the brightness at 1.3~mm is
set by the spectral index $\alpha$, which we leave as a free parameter
that is uniform across the disc. The star is modelled with a template
spectrum corresponding to an effective temperature of 6500~K and a
radius of 1.3~$R_\odot$, which leads to a stellar flux of 30 and 13
$\mu$Jy at 0.88 and 1.3~mm, respectively. In addition, we also fit the
disc inclination $i$ and position angle PA, and nuisance parameters
such as phase center offsets, and the position, size (modelled as a 2D
gaussian) and flux of a SMG found $11\farcs5$ towards the SW of the
star at both frequencies.

While in previous work we left the depth of the gap as a free
parameter, here we opt to leave it fixed after a few tests where we
found that if the depth is not fixed, the location and width of the
gap are not well constrained. Note that this does not mean that we are
forcing the presence of a gap, since the width is allowed to be of
negligible size which is analogous to removing the gap.

In addition to this, we impose a lower limit for $\rmin$ of 14~au,
which is roughly the minimum radius where solids could remain in the
system on stable orbits if HD~206893~B was on a circular orbit at
10~au and had a mass of $\sim12$~\Mjup\ \citep{Wisdom1980,
  Morrison2015}. Given this surface density definition, the disc does
not have a well defined outer edge. Nevertheless, we set a fixed outer
edge of 250~au. This is justified by the absence of significant
emission beyond 250~au and because we do not see any sharp outer edge
in the disc brightness radial profiles. Finally, note that the surface
density could be parametrized differently (e.g. with two belts instead
of one with a gap), however, this choice of parametrization allows to
constrain the gap between the two peaks of emission better using a
single parameter.

\subsection{Results}
\label{sec:results}
\begin{table}
  \centering
  \caption{Best fit parameters of the ALMA data using our parametric
    model. The quoted values correspond to the median, with
    uncertainties based on the 16th and 84th percentiles of the
    marginalised distributions.}
  \label{tab:mcmc}
  \begin{tabular}{lrl} 
    \hline
    \hline
    Parameter & Best fit value & Description\\
    \hline
    $\Md$ [$M_{\earth}$]& $0.031\pm0.002$ & total dust mass \\
    $\rmin$ [au] & $28_{-8}^{+5}$ & disc inner radius \\
    $\rc$ [au]   & $114^{+7}_{-5}$ & disc peak radius \\ 
    $\gamma_1$   & $0.8_{-0.4}^{0+.3} $ & inner disc slope index\\
    $\gamma_2$   & $-3.9^{+0.5}_{-0.6}$ & outer disc slope index\\
    \hline
    $\rgap$ [au] & $74\pm3$ & radius of the gap \\
    $\wgap$ [au] & $27\pm5$& width of the gap\\
    $\dgap$      & 1.0 & fixed fractional gap depth \\
    \hline
    PA [$\degr$] & $61\pm4$& disc position angle \\
    $i$ [$\degr$]& $40\pm3$ & disc inclination from face-on\\
    $\alpha$    &  $2.54\pm0.17$  & disc spectral index \\
    \hline
  \end{tabular}
\end{table}

The best fit parameters are presented in Table~\ref{tab:mcmc}. First,
we find that the disc dust density peaks at
$\rc=114^{+7}_{-5}$~au. Exterior to that the disc surface density
declines steeply as $r^{-3.9\pm0.5}$. Interior to $\rc$, we find that
in order to reconcile the dust levels away from the gap at $\sim30$~au
and 120~au the surface density must increase moderately as
$r^{0.8\pm0.4}$. The disc inner edge is most likely to be around
30~au, although smaller values down to 14~au are still compatible with
the data within $\sim2\sigma$. We can also constrain $\rmin<40$~au at
a 99.7\% confidence level.

In between $\rmin$ and $\rc$, we find that the gap center is well
constrained to $74\pm3$~au. This location is consistent within
$3\sigma$ with the location of the gaps discovered in three other
systems at mm wavelengths \citep{Marino2018hd107, Marino2019,
  MacGregor2019}. The gap width is constrained to $27\pm5$~au. This
constraint on the width can be directly associated to a planet mass,
assuming the gap is truly empty (as assumed in our model to derive its
width) and was cleared by a planet on a circular orbit through
scattering. Roughly speaking, the width of the gap is expected to be
the same as the size of the chaotic zone where mean motion resonances
overlap, i.e. $w_\mathrm{g}\cong 3 \ap \mu^{2/7}$ \citep[where $\ap$
  is the planet semi-major axis and $\mu$ is the planet-star mass
  ratio,][]{Wisdom1980, Morrison2015, Marino2018hd107}. Given the
estimated stellar mass of 1.3~$M_\odot$ \citep{Delorme2017}, the
measured gap width translates to a planet mass of
$0.9^{+0.8}_{-0.5}$~\Mjup, (based on the posterior distribution of
$\rgap$ and $\wgap$), and with 3$\sigma$ upper and lower limits of
3.5~\Mjup\ and 0.03~\Mjup\ (10~\Me), respectively. Therefore the
putative outer planet has a lower mass than the two inner
companions. A caveat in this interpretation is that we have assumed
the gap is empty at its center. If this is not the case and the gap is
not fully empty at its centre, the derived width would be biased
towards smaller values in order to reproduce the same
\textit{equivalent width}. This means that the gap could be shallower
and broader in reality, and more sensitive observations are needed to
assess that. Nevertheless, if the gap was truly carved by a planet, we
expect the surface density of particles to reach zero near the orbit
of the planet \citep[unless there is a large population of solids in
  horseshoe orbits,][]{Marino2018hd107}. Hence fixing the gap depth to
1 is a reasonable assumption and consistent with our interpretation.


The disc orientation is well constrained with $i=40\pm3\degr$ and
PA$=61\pm4\degr$. Note that this inclination is consistent with the
stellar pole inclination \citep[$30\pm5\degr$,][]{Delorme2017}. We
will use the disc orientation later in \S\ref{sec:orbit} as a prior to
fit the orbit of HD~206893~B assuming they are co-planar. The disc
spectral index is found to be $2.53\pm0.16$, i.e. consistent with the
fluxes measured from the images and with other debris discs
\citep{MacGregor2016}.

The parameters to model the SMG that we found within the primary beam
are also well constrained. We find it is centred at a separation of
($-5\farcs6$, $-10\farcs0$), i.e. towards the SW, and has an
integrated flux of 1.5 mJy at 0.88~mm and 0.24 mJy at 1.3~mm. It is
resolved, and its size (or standard deviation) is best fit by a
$0\farcs6x0\farcs3$ 2D Gaussian profile.

\begin{figure}
  \centering \includegraphics[trim=0.0cm 0.0cm 0.0cm 0.0cm, clip=true,
    width=1.0\columnwidth]{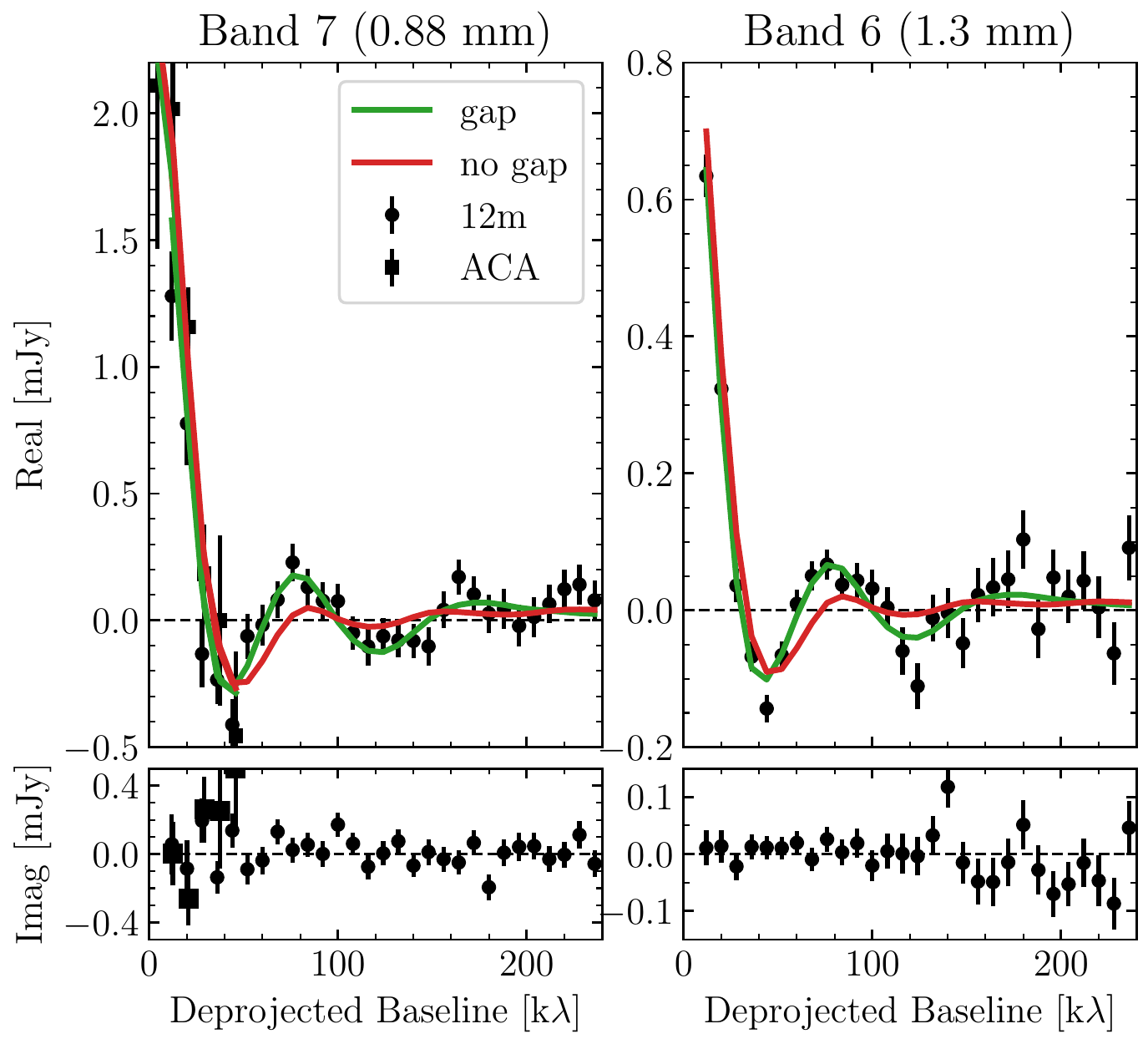}
  \caption{Deprojected visibilities of band 7 (left) and band 6
    (right) data, assuming a disc position angle of $61\degr$ and an
    inclination of $40\degr$. The real and imaginary components of the
    observed visibilities are azimuthally averaged within 8~k$\lambda$
    wide radial bins, and are presented as black error bars in the top
    and bottom panels, respectively. The error bars represent the
    uncertainty estimated as the standard deviation of the observed
    visibilities in each bin divided by the square root of the number
    of independent points. The continuous green and red lines
    represent best fit models with and without a gap, respectively
    (binned using the same procedure as for the data). For better
    display, we crop data points beyond 240~k$\lambda$ since they are
    all consistent with zero.}
 \label{fig:vis}
\end{figure}

Using our best fit model we compute residual maps at the same
resolution as the images in Figure~\ref{fig:alma}, and residual radial
profiles with the same resolution as in Figure~\ref{fig:radial}. We do
not find any residuals stronger than $3\sigma$ within 200~au. To
visualise how well the model fits the data in the visibility space, in
Figure~\ref{fig:vis} we compare the observed deprojected and
azimuthally averaged visibilities with the best fit model (green
line). Note that this procedure assumes the disc is axisymmetric, thus
before averaging we subtract the SMG component from the observed and
model visibilities. Our best fit model reproduces well the multiple
wiggles in the real component of the visibilities, while the imaginary
component is consistent with zero. In the same figure we overlay a
model without a gap for comparison (green line). The model without a
gap fails to reproduce the amplitude of the wiggles beyond
50~k$\lambda$.

\section{Disc orientation and HD~206893~B's orbit}
\label{sec:orbit}
\begin{figure*}
  \centering \includegraphics[trim=0.0cm 0.0cm 0.0cm 0.0cm, clip=true,
    width=0.7\textwidth]{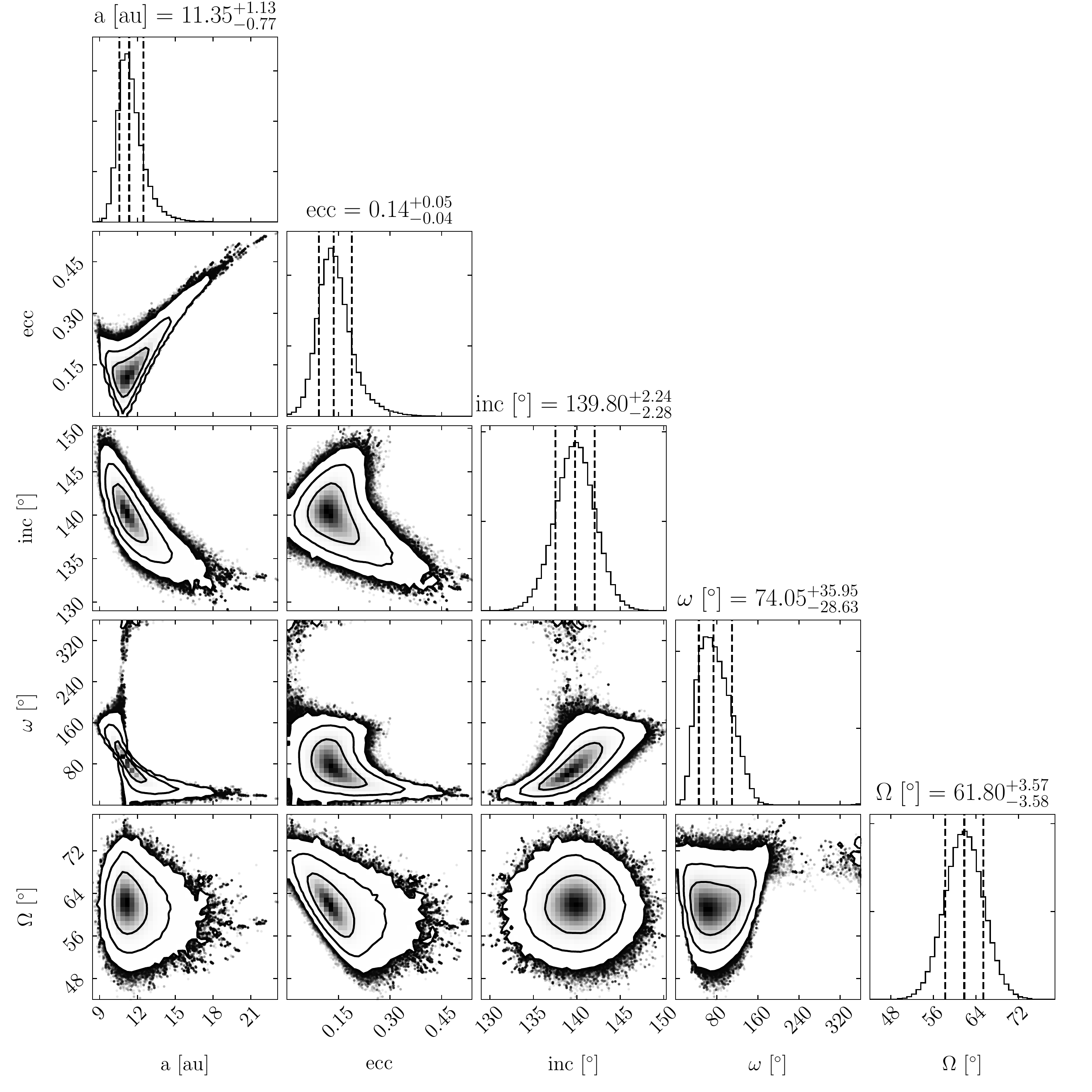}
  \caption{Posterior distribution of the semi-major axis,
    eccentricity, inclination, argument of pericentre and longitude of
    ascending node, derived by fitting the astrometric position of
    HD206893~B using the package \textsc{orbitize} \citep{Blunt2020},
    and using Gaussian priors on the inclination and longitude of
    ascending node. The vertical dashed lines represent the 16th,
    50th, and 84th percentiles. Contours correspond to 68, 95, and
    99.7 per cent confidence regions. This plot was generated using
    the \textsc{python} module \textsc{CORNER} \citep{corner}.}
 \label{fig:corner_orbit}
\end{figure*}

One of the advantages of having tight constraints on the orientation
of the disc is that it can be compared with the orbit of companions
(which could be misaligned), or can be used as prior information when
deriving the orbits of companions assuming both lie in the same
plane. In the case of HD~206893, the orientation of B's orbit is
consistent with the disc orientation derived here
\citep{Grandjean2019}, although the orbit inclination and position
angle are not constrained as well as for the disc. Hence there could
be a moderate misalignment between the two. However, as shown by
\cite{Pearce2014} such misalignments do not last long if the companion
is much more massive than the disc. We expect that in a few secular
timescales the disc will re-orient to the orbital plane of the
companion (if the misalignment is $\lesssim30\degr$). The secular
timescale is only about 10~Myr at 100~au (or shorter if B is more
massive than 12~\Mjup), thus it would be unlikely to observe a
misaligned disc given the age of this system
\citep[50--700~Myr,][]{Delorme2017}.

\begin{figure}
  \centering \includegraphics[trim=0.0cm 0.0cm 16.0cm 0.0cm, clip=true,
    width=1.0\columnwidth]{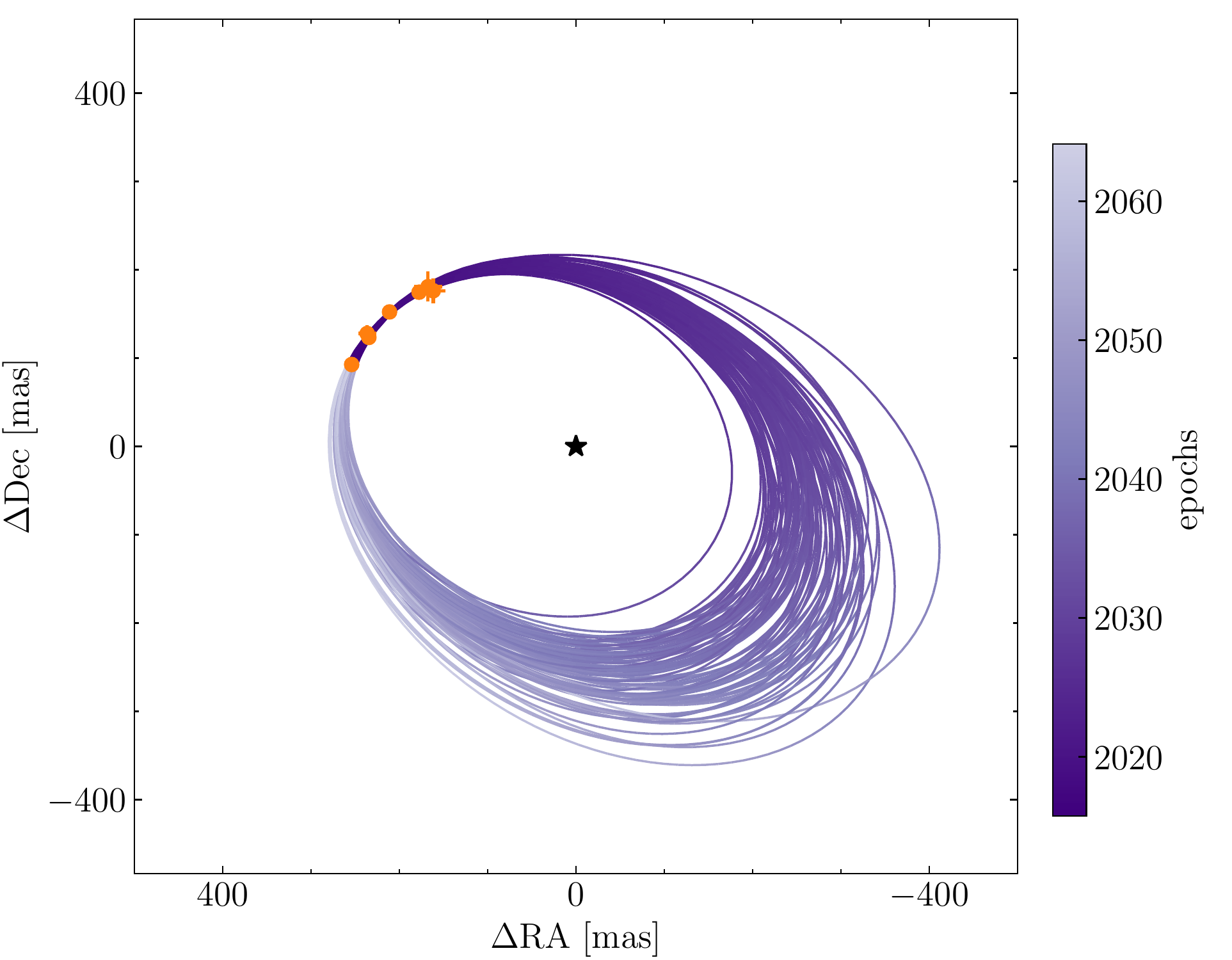}
  \caption{HD206893~B possible orbits, drawn from 100 different points
    of the posterior distribution derived using the package
    \textsc{orbitize} \citep{Blunt2020}. The orange points with
    errorbars represent the astrometric data points.}
 \label{fig:orbits}
\end{figure}

Assuming that the disc and B are co-planar, we can refine its orbital
parameters using as priors the disc inclination and position angle
(longitude of ascending node) derived in \S\ref{sec:results}. To
constrain B's orbital parameters, we use its astrometry reported in
\citet[table 2]{Grandjean2019} and in \citet[table B.1]{Stolker2020},
and the package \textsc{orbitize} \citep{Blunt2020}. This package
allows to recover the posterior distribution of the orbital elements
using a parallel-tempering MCMC algorithm
\citep{Vousden2016}. Additionally, we assume that the SE side of the
disc is the near side, and thus we set normally distributed priors
with an orbital inclination $i\sim\mathcal{N}(140\degr,3\degr)$ and
longitude of ascending node
$\Omega\sim\mathcal{N}(61\degr,4\degr)$. We find that adding the disc
orientation prior breaks some of the degeneracies and reduces the
uncertainties in the estimates. We find B's semi-major axis,
eccentricity and argument of pericentre to be constrained to
$\ap=11.4^{+1.1}_{-0.8}$~au, $\ep=0.14^{+0.05}_{-0.04}$ and
$\omega=74^{+36}_{-29}\degr$, respectively (see Figure
\ref{fig:corner_orbit} for the posterior distribution of the most
relevant parameters using $10^{7}$ points after convergence). These
results show that B's orbit is eccentric (larger than zero with a
$\sim3\sigma$ confidence) if co-planar with the disc, and the
eccentricity is not larger than 0.36 (99.7\% confidence). The
posterior distributions of the inclination and $\Omega$ are centred on
the priors and have uncertainties that are slightly smaller than the
prior distributions. Therefore, there is no indication that the
co-planar assumption is in any tension with the astrometric data. In
Figure~\ref{fig:orbits} we show 100 orbits drawn from the posterior
distribution to illustrate the orientation of the orbit. We find that
the pericentre is more likely to be found in the NW half of the
system, and if very eccentric ($\ep>0.3$) B is currently close to its
pericentre. Using both semi-major axis and eccentricity distributions,
we find a $3\sigma$ upper limit of 22~au on B's apocentre.

Given the estimated eccentricity, we can now assess if B could have
forced an eccentricity in the disc, producing both an offset in the
disc radial structure and a higher disc flux near apocentre
\citep[i.e. apocentre glow,][]{Pan2016}. Since in \S\ref{sec:radial}
we found that the NE half of the disc was brighter, it is fair to
assume $\omega\sim180\degr$ and thus $\ep\lesssim0.3$ (i.e. we assume
the planet and disc are apsidally aligned which is valid if the
companion is much more massive than the disc). This means that the
forced eccentricity, $e_\mathrm{f}=5a_\mathrm{BD}e_\mathrm{BD}/(4a)$
\citep{MurrayDermott1999, Mustill2009}, imposed by the BD on particles
with semi-major axis $a$ near the disc peak density ($\rc$) is
$\lesssim0.03$. Such a low forced eccentricity would not produce a
detectable contrast in brightness between apocentre and pericentre
\citep[$\sim1.03$,][]{Pan2016}. For any other value of $\omega$, we
still expect low forced eccentricities ($\lesssim0.06$), mainly due to
the small ratio between B's semi-major axis and the disc distance, and
thus not detectable asymmetries due to interactions with B alone.

\section{Discussion}
\label{sec:dis}

In this section we summarise and discuss the different dynamical
constraints on companions in HD~206893 (\S\ref{sec:3bodies}), and we
discuss the different potential origins of the gap (\S\ref{sec:gap}),
disc stirring (\S\ref{sec:stirring}) and the ubiquity of gaps in
exoKuiper belts (\S\ref{sec:gaps}).

\subsection{3 companions}
\label{sec:3bodies}

In this section we try to put in context the different observational
constraints on HD~206893 to conclude on some basic properties of its
companions. First, by combining the different astrometric data of
HD~206893~B and the derived disc orientation in this work, we
constrain the orbit of B to a semi-major axis of
$11.4^{+1.1}_{-0.8}$~au and an eccentricity of
$0.14^{+0.05}_{-0.04}$. Its mass remains uncertain in the range
12--50~\Mjup, given its uncertain age
\citep[50--700~Myr,][]{Delorme2017}, which also means that B could
have truncated the disc through scattering or via secular resonances
(see \S\ref{sec:secularres}). Since no additional companions have been
detected with direct imaging beyond $\sim4$~au, we can use this as an
upper limit on the magnitude or mass of additional companions. We use
the $5\sigma$ detection limits from \cite{Milli2017hd206, Delorme2017}
and transform these to planet masses using AMES-COND models
\citep{Baraffe2003}. Assuming an age of 50~Myr (700~Myr) we can rule
out the presence of planets more massive than
$\sim3$~\Mjup\ ($\sim10$~\Mjup) from 4 to 160 au.

Second, this system has a proper motion anomaly of
$96\pm28$~m~s$^{-1}$ when subtracting the proper motion measured by
Gaia DR2 from the one estimated using DR2 and Hipparcos astrometry,
which has a longer baseline \citep{Kervella2019}. Although its
magnitude is consistent with the dynamical effect that B would have on
the star, as shown by \cite{Grandjean2019} the proper motion anomaly
(PMa) has a position angle ($233\pm12\degr$) that is inconsistent with
the location of B (PA$\sim70\degr$ in 2015.5). This means that there
is likely an additional inner companion (HD~206893~C), which would be
also responsible for the RV drift of
$87^{+16}_{-14}$~m~s$^{-1}$~yr$^{-1}$ detected by
\cite{Grandjean2019}. Note that B is not sufficiently massive or close
to the star to explain this RV acceleration. Given the inclination of
this system ($40\degr$) and the length of this trend, C should have a
semi-major axis larger than 1.4~au and a mass greater than
$\sim10$~\Mjup\ to produce an acceleration of
$\sim90$~m~s$^{-1}$~yr$^{-1}$. This inner companion would dominate the
observed stellar velocity around the center of mass ($v_\star\propto
m/\sqrt{a}$, with $m$ and $a$ the mass and semi-major axis of a
companion) and thus the PA of the PMa is not expected to be correlated
with the position of B. Using N-body simulations with the
\textsc{python} package \textsc{REBOUND} \citep{rebound, mercurius} we
confirm this and find that both direction and magnitude of the PMa can
be explained by C, even in the presence of B further out.



Third, the gap in the planetesimal belt centered at $74\pm3$~au and
extending $27\pm5$~au suggests the presence of a third companion
(HD~206893~b) that carved this gap around its orbit. The gap's width
and center constrain the planet mass to $0.9^{+0.8}_{-0.5}$~\Mjup\ and
semi-major axis to $74\pm3$~au (based on the posterior distribution of
the gap width in \S\ref{sec:results}). Such a planet would have
remained undetected in the existing direct imaging data. The SPHERE H2
observations reached a contrast of 16~mag at that distance, whereas
AMES-COND models predict a contrast of 18 (27) mag if 50 (700) Myr
old. Although detecting b is beyond the current capabilities of direct
imaging instruments, it will certainly be within reach of JWST with
NIRCam or MIRI at longer wavelengths. Note that these constraints
assume that the gap was indeed carved by a planet in situ clearing its
orbit through scattering. In \S\ref{sec:gap} we discuss this
possibility and alternative scenarios.

\begin{figure}
  \centering \includegraphics[trim=0.0cm 0.0cm 0.0cm 0.0cm, clip=true,
    width=1.0\columnwidth]{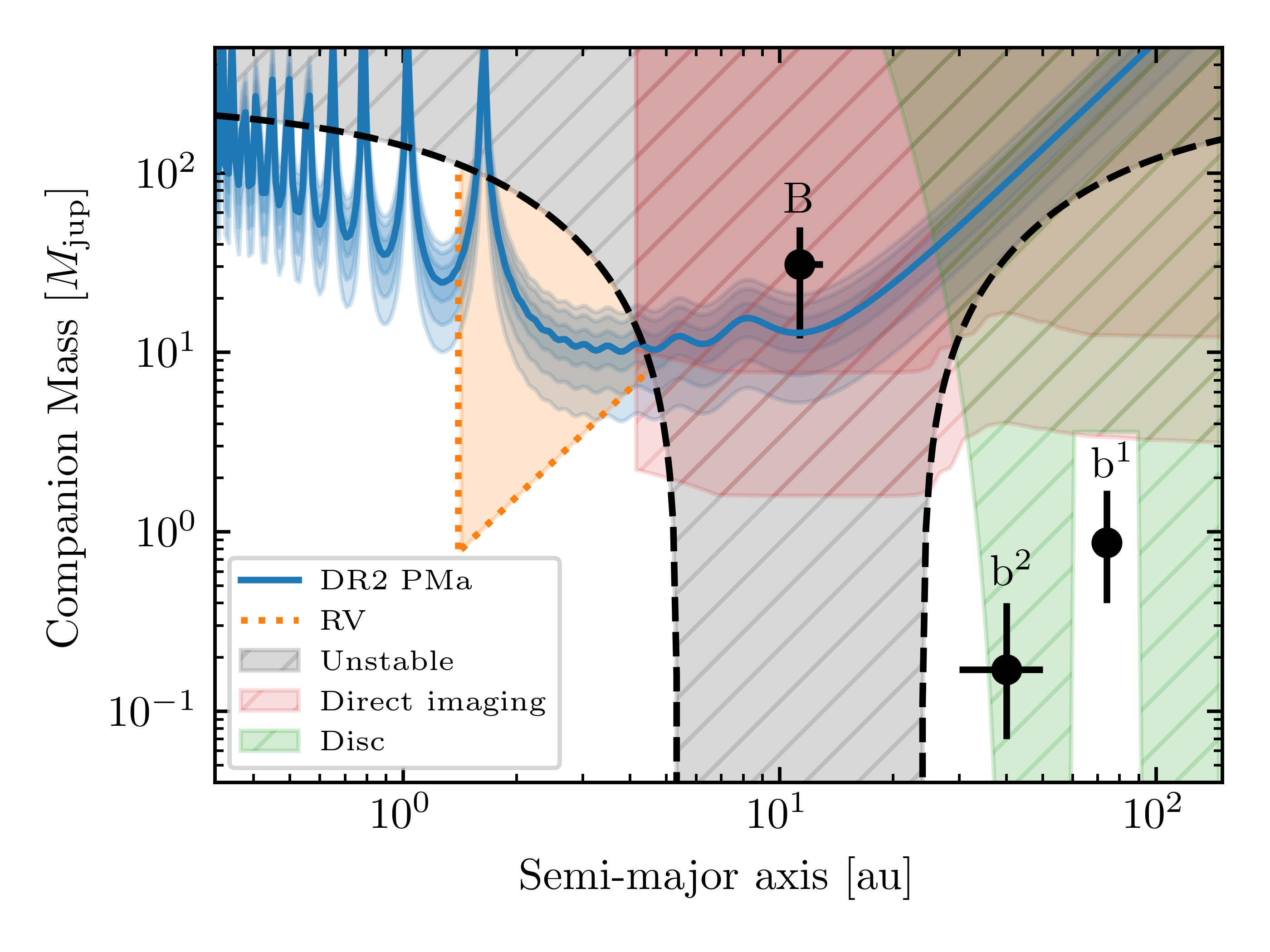}
  \caption{Companion mass vs semi-major axis in HD~206893. The black
    errorbars represent HD~206893~B and two possible masses and
    semi-major axes of the putative planet b according to two
    different scenarios (see Table~\ref{tab:planets}). The blue line
    and shaded region (68\%, 95\% and 99.7\% confidence levels)
    represent the companion mass of an additional companion (C),
    necessary to explain the PM anomaly, calculated using the Gaia DR2
    proper motion anomaly. The black dashed line represents the mass
    above which the orbit of a companion would be unstable based on a
    5 mutual Hill radii criteria, and thus companions are excluded in
    the grey hatched region. The orange dotted lines represent the
    minimum semi-major axis and minimum mass of a companion to explain
    the RV drift. Taking this consideration and the stability limits,
    the companion responsible for the RV drift must lie in the orange
    shaded region. The red hatched regions represent the companion
    masses that would have been detected with SPHERE in addition to B
    assuming ages of 700 Myr (top) and 50 Myr (bottom). The green
    hatched region represents the companion masses and semi-major axes
    that are ruled-out by the presence of the disc.}
 \label{fig:Mvsa}
\end{figure}

In Figure~\ref{fig:Mvsa} and Table~\ref{tab:planets} we summarise all
these constraints on companions around HD~206893. The black points
with errorbars represent HD~206893~B and the putative planet b if the
gap was cleared through scattering (labelled as 1, see
\S\ref{sec:p74}) or if it was cleared through secular interactions
(labelled as 2, see \S\ref{sec:psec}). The red hatched regions
represent $5\sigma$ upper limits from direct imaging assuming ages of
50 (bottom) and 700~Myr (top). The green hatched region shows the
planets that are ruled out if on a circular orbit since they would
either push the disc inner edge beyond 40 au (inconsistent with our
observations), produce a gap at 74~au much wider than observed, or
carve additional gaps in the disc that are not observed. Note that in
the secular interaction scenario the planet can be located in a region
where the disc density is high \citep{Pearce2015doublering}. The blue
line and shaded region (68\%, 95\% and 99.7\% confidence levels)
represent the mass of C in order to explain the magnitude of
HD~206893's PMa. This mass is calculated using the Gaia DR2 proper
motion anomaly (after subtracting the PM from Hipparcos-Gaia DR2),
equations in \cite{Kervella2019}, and the disc orientation derived in
this work. The confidence levels were determined using Monte Carlo
simulations (bootstrapping the PMa and disc orientation). In addition,
the dashed black line represents the mass above which the system would
become unstable in a short timescale, here simply defined as an
orbital spacing smaller than 5 mutual Hill radii assuming B is on a
circular orbit and has a mass of 12~\Mjup. Thus, we can exclude
companions in the grey hatched region. Note that the stability
criteria depends on several factors which we are not exploring here
and thus this upper limit should be taken with caution \citep[see
  details in][]{Smith2009stability}. Moreover, we found that B is
likely on an eccentric orbit and thus the allowed planet masses would
be even lower \citep{Gladman1993, Lazzoni2018}. In the same figure we
overlay the RV constraints on C, namely its minimum semi-major axis
(vertical orange line) and minimum mass (diagonal orange line)
calculated to produce an RV acceleration of at least
45~m~s$^{-1}$~yr$^{-1}$ \citep[$3\sigma$ lower
  limit,][]{Grandjean2019}. The orange shaded region therefore
represents a conservative range of companions that could explain
the RV trend and still be stable. When combined, these constraints
imply that the innermost companion should lie in the intersection
between the orange wedge and the blue shaded area to explain the RV
trend and proper motion anomaly, i.e. have a semi-major axis of
roughly 1.4--4.5~au and a mass in the range 4--100~\Mjup. This massive
inner companion would lie at a projected separation of 26--110~mas,
and thus it could be resolved using sparse aperture masking on SPHERE
or with GRAVITY at K band using molecular mapping. It is difficult to
predict its exact PA since its period is not well constrained
(1.5--8.4 years) and is comparable to the period over which Gaia
obtained astrometric observations \citep[1.8 years,][]{Gaiadr2}. Hence
the intrinsic velocity vector of the star has been significantly
averaged over time \citep{Kervella2019}, and thus its direction is
biased. The analysis presented here illustrates how the combination of
NIR direct imagining, ALMA imaging, RV and PMa constraints can be used
in combination to tightly constrain the architecture of a system.


Note that while we do not have strong constraints on the
eccentricities of HD~206893~b and C, if B is truly eccentric we expect
that the two additional companions will gain eccentricity through
secular interactions even if they had initially zero
eccentricities. Over Myr timescales they will exchange eccentricities
and thus b and C could have eccentricities as high as B depending on
their masses and semi-major axes.

\begin{table}
  \centering
  \begin{adjustbox}{max width=1.0\columnwidth}
    \begin{threeparttable}
      \caption[]{Summary of the inferred properties of the three
        companions in the HD~206893 system based on direct imaging,
        RV, proper motion anomaly and disc imaging with ALMA.}
      \label{tab:planets}
      \begin{tabular}{lcccl} 
      \hline
      \hline
      Designation & a & e & Mass & reference  \\
                  &  [au]& & [\Mjup]  & \\
      \hline
      HD 206893 (AC)B  &$11.2^{+1.5}_{-0.9}$ & $0.14^{+0.07}_{-0.06}$ & $12-50$ & (1, 2, \S\ref{sec:orbit})\\
      HD 206893 (A)C &$1.4-4.5$ & - &  $4-100$ & (1, 3, \S\ref{sec:3bodies})\\
      HD 206893 (ABC)b$^{1}$ &$74^{+3}_{-3}$ & - & $0.9^{+0.8}_{-0.5}$ & (1, \S\ref{sec:p74}) \\
      HD 206893 (ABC)b$^{2}$ &$30-50$ & - & 0.07--0.4 & (1, \S\ref{sec:psec}) \\
      \hline
    \end{tabular}
    \begin{tablenotes}
    \item \textbf{References used in this table:} (1) this work; (2)
      \cite{Delorme2017}; (3) \cite{Grandjean2019}. $^1$: putative
      planet opening the gap around its orbit through
      scattering. $^2$: putative planet opening the gap beyond its
      orbit through secular interactions. Its semi-major axis is
      expected to be at the disc inner peak and its mass equal to the
      total disc mass.
    \end{tablenotes}
    \end{threeparttable}
  \end{adjustbox}
\end{table}

\subsection{The gap origin}
\label{sec:gap}
In this section we discuss potential dynamical origins for the
observed gap based on different scenarios explored in previous work
and the multiple information that exists for this system, namely the
brown dwarf detected at a 11~au separation and the inner companion
responsible for the RV drift and proper motion anomaly (see
\S~\ref{sec:3bodies}).

\subsubsection{Gap clearing by planet at 74 au}
\label{sec:p74}

As commented in \S\ref{sec:results}, the gap discovered at 74~au hints
at the presence of a planet on a circular orbit clearing the material
near its orbit. The putative planet would need to be a gas or ice
giant of mass $0.9^{+0.8}_{-0.5}$~\Mjup (based on the posterior
distribution of the gap width), in order to create a $27\pm5$~au wide
gap via scattering of particles within its chaotic zone. Such a
planet, if massive enough to retain a significant atmosphere, probably
formed while a gas rich protoplanetary disc was still present and
possibly before the planetesimal belt was formed. In that hypothetical
case, the distribution of solids would have been already truncated
around the planet's orbit. As shown in \cite{Marino2019}, if the gap
was inherited from a gap in the distribution of dust which then grew
to form the planetesimal belt, then the putative planet could have a
lower mass of a few tens of Earth masses. In fact, several gaps are
observed at those distances in protoplanetary discs
\citep[e.g.][]{Andrews2018, Huang2018, Long2018}, hence it is possible
that the gap could have been inherited from the dust distribution in
HD~206893's protoplanetary disc. While growing, this planet could have
also migrated through the gas rich disc, or via planetesimal driven
migration after disc dispersal possibly widening the gap. This means
that the required planet mass to open such a gap could be much
lower. Therefore, the derived planet masses must be taken with
caution. Nevertheless, the gap width does impose a strict upper limit
of 3.5~\Mjup, otherwise the gap width would be much greater than the
ALMA observations show.

If the putative planet was on a slightly eccentric orbit, it would
likely clear an asymmetric and eccentric gap and force an eccentricity
in the disc \citep[e.g.][]{Pearce2014}. Note that this putative planet
could become eccentric simply through secular interactions with B
which is likely on a eccentric orbit (\S\ref{sec:orbit}). This
eccentricity could then explain the tentative $30\pm13\%$ brightness
asymmetry. Detailed N-body simulations that explore the level of disc
asymmetry produced by a planet in the middle of the disc with
different eccentricities are needed to fully assess this scenario.

While a planet with an orbit within the gap is plausible, this
scenario triggers many questions since the gaps found around
HD~107146, HD~92945 and HD~15115
\citep{Marino2018hd107,Marino2019,MacGregor2019} lie all at a very
similar radius (59--74~au). Naively we would expect these radii to
differ by more given the multiple factors that determine a planet's
final orbit and the wide range of radii covered by these discs
(40--150~au). This is not an issue that applies only to this scenario,
but rather to any dynamical mechanism that requires the presence of
planets and fine tuning of their parameters. Alternatively, the gaps
could lie at a similar radius if they are formed instead as a
consequence of a change in planetesimal properties (e.g.  inefficient
planetesimal formation, or weaker planetesimals that collisionally
evolve faster) or a hotspot for efficient planet formation. These
questions are difficult to answer with a limited sample of discs
observed with enough sensitivity and resolution to detect these gaps
(see \S\ref{sec:gaps}). A future ALMA survey could expand this sample
and constrain the ubiquity of gaps and determine their radius and
width distribution.

\subsubsection{Secular resonances between planet at 2 au and BD at 11 au}
\label{sec:secularres}

\begin{figure}
  \centering \includegraphics[trim=0.0cm 0.0cm 0.0cm 0.0cm, clip=true,
    width=1.0\columnwidth]{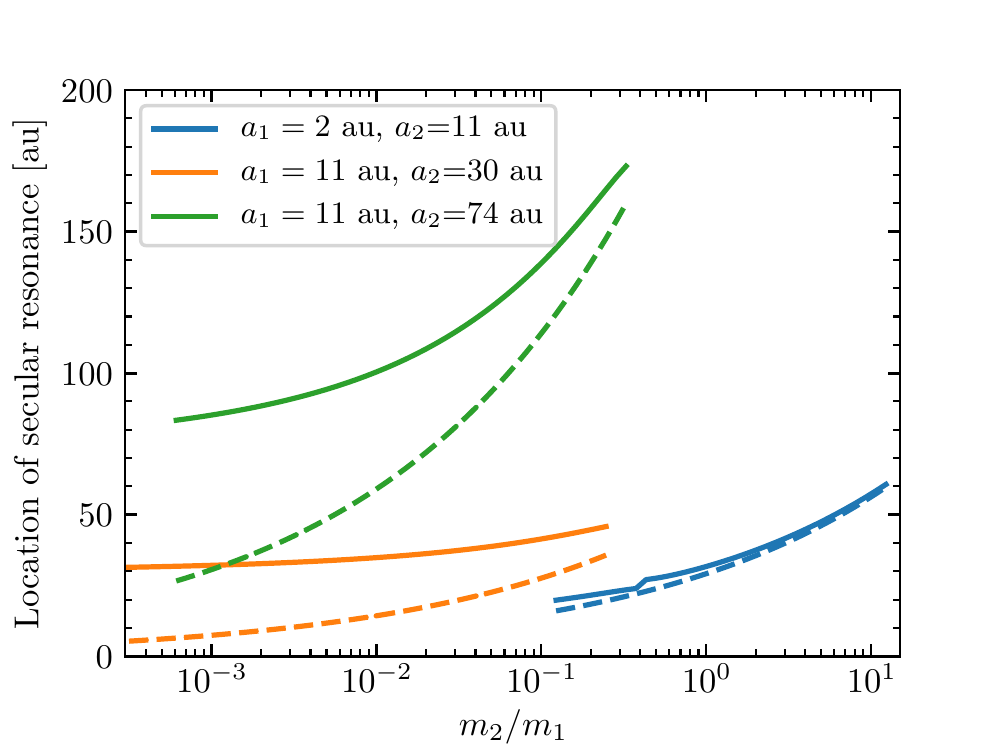}
  \caption{Location of the strongest secular resonance exterior to the
    orbits of different pairs of companions: massive inner companions
    at 2 and 11 au (blue), brown dwarf at 11 au and hypothetical
    planet at 30~au (orange), brown dwarf at 11 au and putative planet
    in the gap at 74 au (green). The continuous line represent the
    true location of the resonance, while the dashed line its location
    approximated by Equation \ref{eq:SR}. Only mass ratios consistent
    with the dynamical constraints are displayed. The discontinuity on
    the line is due to one of the two exterior resonances becoming
    stronger than the other above a mass ratio $\sqrt{a_1/a_2}$, while
    the other becomes negligible (we only plot the strongest). $m_1$
    and $m_2$ stand for the mass of the inner and outer companion,
    respectively. }
 \label{fig:SR}
\end{figure}

Given the two companions on orbits interior to the disc inner edge,
this system is well suited to test if secular resonances exterior to
their orbits, could create a gap in the disc at 74~au where particles
eccentricities are excited to large values \citep{Yelverton2018}. To
assess this, we solve Equation 13 in
\cite{Yelverton2018}\footnote{There is a typo in Equation 13 in
  \cite{Yelverton2018}. In the right hand side, the first factor
  should be $a_1^{1/2} a_2^2 a^{-5/2}$ instead of
  $(a_1/a_2)^{1/2}$. We confirmed this through private communication
  with Yelverton et al.} to find the gap/resonance location using the
estimated masses and semi-major axes of the two inner companions. We
find that secular resonances cannot open a gap at such a large
distance with respect to HD~206893~B (see solid blue line in
Figure~\ref{fig:SR}). Since the mass ratio between B and C is
$\lesssim10$ (see Table~\ref{tab:planets}), we find that the strongest
secular resonance would be located at $\lesssim50$~au. Therefore this
scenario cannot explain the gap at 74~au.

Nevertheless, secular resonances could have instead truncated the disc
inner edge. Under the influence of a single planet on an eccentric
orbit, \cite{Pearce2014} showed that the disc should be truncated at a
semi-major axis (at apocentre)
\begin{equation}
  a_\mathrm{in}\approx \ap(1+\ep)+5 R_\mathrm{H,Q},
\end{equation}
where $R_\mathrm{H,Q}$ is the Hill radius of an eccentric planet at
apocentre. Under the influence of two planets, that inner edge could
be further out if the strongest secular resonance is located beyond
$a_\mathrm{in}$. Using equation 24 in \citet[][]{Yelverton2018} we
find the approximate location of the resonance to be (dashed blue line
in Figure \ref{fig:SR})
\begin{equation}
r^{\mathrm{SR}}_{\min}= 29
\left(\frac{a_{2}}{11\ \mathrm{au}}\right)^{11/7}
\left(\frac{a_{1}}{2\ \mathrm{au}}\right)^{-4/7} \left(\frac{m_{2}
}{m_{1}}\right)^{2/7} \ \mathrm{au}, \label{eq:SR}
\end{equation}
where $a_{1}$ and $a_{2}$ are the semi-major axes of the companions,
while $m_{1}$ and $m_{2}$ are their masses, with 1 designating the
innermost companion and 2 the BD. In order to effectively truncate the
disc, the width of the secular resonance would need to be large enough
to excite solids from $a_\mathrm{in}$ to
$r^{\mathrm{SR}}_{\min}$. This width is controlled primarily by the
eccentricity of HD~206893~B. Given the uncertainties on its
eccentricity ($0.14^{+0.05}_{-0.04}$) and its mass \citep[$12-50$
  \Mjup,][]{Delorme2017} it is hard to assess whether secular
resonances will deplete a region that otherwise would be stable.
Assuming the best fit values of our fit and using equation 36 in
\cite{Yelverton2018} we expect a secular resonance width of
$\sim20$~au. Therefore it is possible that the disc inner edge is not
truncated by the BD, but by secular resonances. Note that HD~206893~B
alone could truncate the disc out to $\sim30$~au if it has an
eccentricity of 0.2 and a mass of 50~\Mjup. A more precise
characterisation of HD~206893~B's orbit and the disc inner edge could
provide better estimates of the masses of the two inner companions.

\subsubsection{Secular resonances between BD and planet at 30 au}

Secular resonances might still have created the gap if instead the
resonance at 74 au was with the BD at 11~au and an outer planet
sitting just interior to the disc inner edge at around 30~au. We find
that this putative planet and the brown dwarf at 11 au would need to
have a mass ratio close to unity (see solid orange line in
Figure~\ref{fig:SR}). This is ruled-out by direct imaging which did
not detect any companion more massive than 3~\Mjup\ at 30~au
\citep[assuming an age of 50~Myr,][]{Delorme2017}.


\subsubsection{Secular resonances between BD and planet at 74 au}

If the gap at 74~au was indeed carved by a planet at that distance, it
is worth discussing what other observables could confirm this
scenario. Given the range of possible masses of the BD and the
putative planet c at 74~au, a secular resonance could be located in
between the orbit of b and the disc outer edge. We find that the
resonance would be located between 85 and 170~au for mass ratios of b
and B of $10^{-3}-0.3$ (solid green line in Figure~\ref{fig:SR}). This
resonance would create a gap that is expected to be very wide
\citep[$\gtrsim50$~au if b has an eccentricity $\gtrsim0.05$ according
  to Equation 36 in][]{Yelverton2018}, and thus noticeable by our
observations. This means that we can already rule out that this gap is
present in between 74 and 110~au, otherwise the gap would be seen to
be much wider and the disc peak further out (or at 40~au
instead). Therefore, imposing that the secular resonance is located
beyond 120~au, we can constrain the mass ratio to be larger than 0.04
(consistent with the expected range, see
Table~\ref{tab:planets}). This means the putative planet would need to
be at least 0.5~\Mjup\ in mass, and thus likely a gas giant.

There could already be evidence of the presence of this secular
resonance in between 120 and 170~au. Such a resonance could be
responsible for shaping the distribution of solids in the outer
regions, giving HD~206893's disc its observed appearance. As figures
\ref{fig:radial} and \ref{fig:comparison} show, its outer edge is not
sharp or well defined. The surface brightness declines smoothly with
radius down to the noise level at 180~au. The smooth decline could be
due to solids in high eccentricity orbits, similar to a scattered disc
\citep{Booth2009, Wyatt2010, Marino2017etacorvi, Geiler2019}. In
contrast, HD~107146 and HD~92945 have well defined sharp outer
edges. Therefore, this smoother outer edge in HD~206893 could be due
to the effect of secular resonances exciting the eccentricities of
particles over a broad region in the outer edge of this disc.

A caveat in the reasoning above is that the analysis in
\cite{Yelverton2018} is only strictly valid for a system with only two
planets and a massless disc (or with a mass much smaller than the
planet masses). Taking into account the disc mass and the effect of
the innermost companion is beyond the scope of this paper, but we
acknowledge that it is important for the location of secular
resonances and thus is crucial for constraining the mass of this
putative planet b.



\subsubsection{Secular interactions between the disc and a scattered planet}
\label{sec:psec}
A different possibility to explain the observed gap and tentative disc
asymmetry is through secular interactions between the disc and a
planet of a similar mass \citep{Pearce2015doublering}. In this
scenario a planet is scattered out by a more massive one (e.g. B) onto
an eccentric orbit that overlaps with the disc. Through secular
interactions the low mass planet is able to open a broad and
asymmetric gap in the disc, which could explain our observations. For
this to happen, the disc must be wide and its mass must be comparable
to the planet mass. Although it is hard to assess the total disc mass
since observations are not sensitive to planetesimals, we use the
collisional model described in \cite{Marino201761vir} and estimate the
total disc mass to be around $\sim20-120$~\Me\ assuming the disc has
been collisionally evolving for 50-700~Myr and planetesimals have a
maximum size of 100~km. Therefore, we can already rule-out that the BD
opened the observed gap since it is at least 40 times more massive
than the disc. Moreover, the model by \cite{Pearce2015doublering}
predicts that HD~206893~B would be located just interior to the gap
where the disc surface density peaks ($30-50$~au according to
Figure~\ref{fig:radial}), which is inconsistent with B's position at
11 au.

Nevertheless, B could have been the massive companion that scattered
out an additional less massive planet which could have opened the gap
and would now reside in between the disc inner edge and the gap. Note
that this scenario proposed by \cite{Pearce2015doublering} did not
take into account the influence of an inner massive companion and thus
it is unclear how exactly its presence affects the secular evolution
and gap opening. Simulations tailored to this system could help to
assess in detail if this scenario could be at play. The last row of
Table~\ref{tab:planets} shows the mass and orbital constraints on this
putative planet for this scenario.





\subsubsection{Planet-less scenarios}

Finally, we consider whether the gap could have been opened without
the influence of a planet. As mentioned in \cite{Marino2019}, gaps in
the distribution of mm-sized grains could be due to changes in
planetesimal properties within the gap (e.g. strength, maximum size,
porosity). Such changes could change the size distribution and
collisional evolution in a way that it produces a depletion on
mm-sized grains. It is uncertain how large those changes would need to
be and if this is something plausible according to planetesimal
formation models.

Another possibility is that dust-gas interactions could lead to gaps
or ring-like structure \citep[e.g. through photoelectric
  instabilities][]{Klahr2005, Lyra2013photoelectric, Richert2017}. As
in \cite{Marino2018hd107, Marino2019}, we can also rule out that for
this system gas drag could be important for mm-sized grains. Using our
CO mass upper limit (\S\ref{sec:gas}) and dust mass
(\S\ref{sec:results}) we find that the CO gas-to-dust mass ratio is at
least $10^{-4}$ and thus even if H$_2$ was present (with a typical
ISM-like H$_2$/CO abundance of of $10^{4}$) we expect a dust to gas
mass ratio much smaller than unity. Moreover, collisional timescales
are orders of magnitude shorter than stopping times due to gas drag,
and thus its effect is expected to be negligible for mm-sized grains.

A third possibility is that the gaps were inherited from the solid
distribution in protoplanetary discs, where gaps in the dust
distribution can be created via different methods without the
intervention of any planet. For example, dead zones
\citep{Pinilla2016}, MHD zonal flows \citep{Flock2015}, secular
gravitational instability \citep{Takahashi2014}, instabilities
originating from dust settling \citep{Loren-Aguilar2015}, dust
particle growth by condensation near ice lines \citep{Saito2011}, and
viscous ring instability driven by dust \citep{Dullemond2018}. All
these mechanisms can shape the distribution of solids in
protoplanetary discs, which could be later inherited by planetesimals
formed from those solids, although how exactly this would be inherited
is uncertain given the unconstrained planetesimal formation
process. If any of those models were to predict a fixed gap radius
around 70~au, then this could explain the distribution of gaps
observed so far in exoKuiper belts.

\subsection{Disc stirring}
\label{sec:stirring}
Since the observed dust must be replenished through collisions of
larger planetesimals \citep[e.g.][]{Wyatt2002}, their orbits must be
stirred above a certain level such that relative velocities are high
enough to cause destructive collisions. Planetesimals are expected to
have nearly circular orbits when protoplanetary discs disperse,
therefore stirring should take place after disc dispersal and on a
timescale shorter than the age of the observed system. There are two
main stirring mechanisms proposed. Planetesimals could be stirred by
planets or more massive companions
\citep[e.g.][]{Mustill2009}. Alternatively, big planetesimals within
the disc could stir it and ignite a collisional cascade
\citep[i.e. self-stirring,][]{Kenyon2008, Kenyon2010, Kennedy2010,
  Krivov2018stirring}.

\begin{figure}
  \centering \includegraphics[trim=0.0cm 0.0cm 0.0cm 0.0cm, clip=true,
    width=1.0\columnwidth]{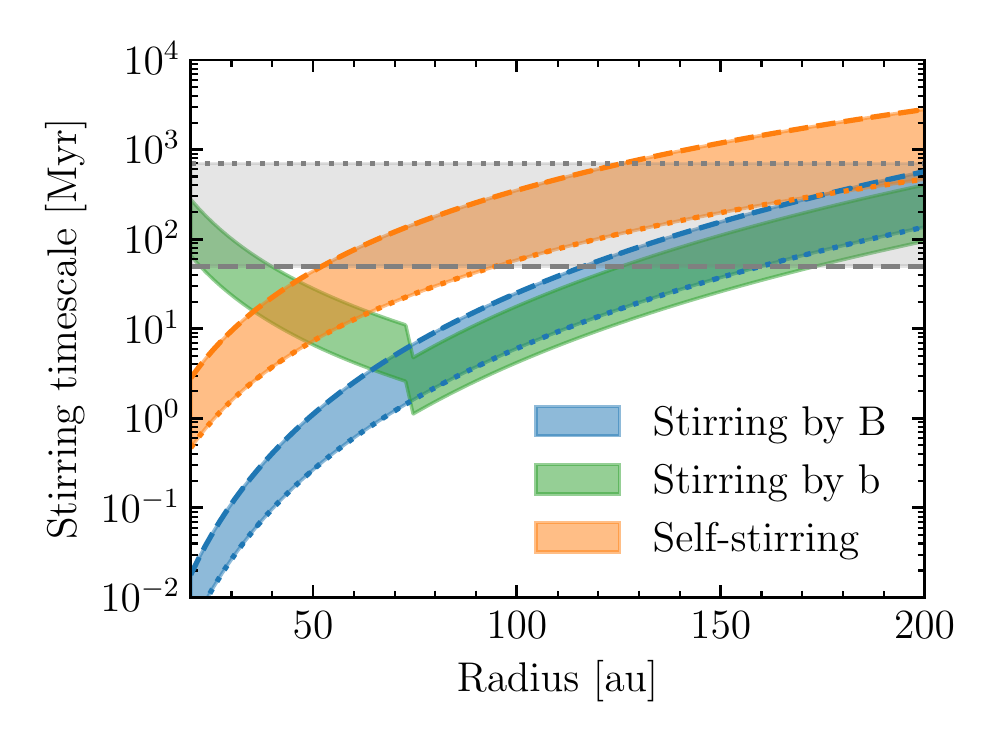}
  \caption{Comparison between the age of HD~206893 (grey region) and
    stirring timescales due to HD~206893~B (blue), b (green) and
    self-stirring (orange) as a function of disc semi major axis or
    radius. The width of the shaded regions represent the respective
    uncertainties due to the age uncertainty, which could range
    between 50 (dashed line) and 700~Myr (dotted line).}
 \label{fig:stirring}
\end{figure}

For HD~206893 we know the system has a massive companion at 11 au
which may have stirred the disc. Based on its estimated eccentricity
of $0.14^{+0.05}_{-0.04}$ (\S\ref{sec:orbit}), we find it could stir
solids to eccentricities above 0.01 out to 150~au \citep[equation 8
  in][]{Mustill2009}, unless its eccentricity is below 0.1. In Figure
\ref{fig:stirring} we compare the age of the system with the timescale
it would take to stir solids due to secular interactions with B
\citep[equation 15 in][]{Mustill2009} for a mass in the range
12--50~\Mjup\ (blue shaded region). Curves corresponding to a system
age of 50 (700) Myr are represented with dashed (dotted) lines. We
find that B could have stirred the disc out to $\sim100$~au if the
system is young, or beyond 200~au if it is rather old. Hence, unless B
has a low eccentricity or the system is young, B alone could explain
that the disc is stirred at least out to 150~au. Moreover, if the
putative planet b is indeed at 74~au, has a mass between
0.4-1.7~\Mjup\, and is on a mildly eccentric orbit (0.02), we find
that it could have stir the disc even on a shorter timescale than B
beyond 74~au \citep[green shaded region; the discontinuity around 74
  au is due to the use of equation 16 in][for
  $r<74$~au]{Mustill2009}. Note that the innermost companion would be
very inefficient at stirring solids at large radii given its small
semi-major axis.

It is also worth considering self-stirring to see if both mechanisms
could be at play. Using Equation 33 in \cite{Krivov2018stirring}, a
maximum planetesimal diameter of 100~km and a disc mass of
20--120~\Me, we find self-stirring is too slow to excite
eccentricities beyond 100~au. Considering a larger maximum
planetesimal diameter of 400~km, and correspondingly a disc mass
larger by a factor $\sim1.3$ to fit the same surface density of dust
\citep[see equation 7 in][]{Marino2019}, we find that the disc could
have been self-stirred out to 150~au if older than 500~Myr (orange
shaded region). Nevertheless, the timescale for self-stirring is
overall longer than stirring by B or b, therefore we conclude that the
disc was likely stirred by the companions if born unstirred \citep[a
  similar conclusion was reached by][for the debris disc around
  HD~193571]{MussoBarcucci2019}.

\subsection{Are gaps common among exoKuiper belts?}
\label{sec:gaps}

Based on the literature, there are only 6 exoKuiper belts that so far
have been observed with ALMA with enough resolution and sensitivity to
detect gaps (e.g. $\gtrsim4$ beams across their width and a S/N
$\gtrsim10$ in the deprojected radial profile). These are HD~107146,
HD~92945, $\beta$~Pic, HD~15115, AU~Mic \citep{Marino2018hd107,
  Marino2019,Matra2019betapic, MacGregor2019, Daley2019}, and now
HD~206893. Four of these show evidence of gaps (HD~107146, HD~92945,
HD~15115 and HD~206893), suggesting gaps could be common in exoKuiper
belts, at least among wide and bright discs. This number could grow to
five if we consider AU~Mic's best fit model of a narrow inner ring in
addition to a broader outer disc, as evidence of a gap in between
these two components. Such ubiquity is not rare among protoplanetary
discs, with the majority of discs larger than 50~au showing radial
substructure in the form of gaps and rings \citep[e.g][]{Andrews2018,
  Long2018, Long2019}. This means gaps could be truly ubiquitous in
both large protoplanetary and planetesimal belts. Whether this type of
substructure can be directly linked between the two, e.g. gaps in
planetesimal belts could be inherited from the dust distribution in
protoplanetary discs, is still unclear. Future ALMA observations of a
large sample of debris disc could reveal the distribution of gap radii
and widths, and thus allow for a statistical comparison between the
two. Moreover, gaps could be also present in narrow exoKuiper belts as
well, as it is in some protoplanetary discs
\citep[e.g. HD~169142,][]{Perez2019}, and this is something yet to
explore.

\begin{figure}
  \centering \includegraphics[trim=0.0cm 0.0cm 0.0cm 0.0cm, clip=true,
    width=1.0\columnwidth]{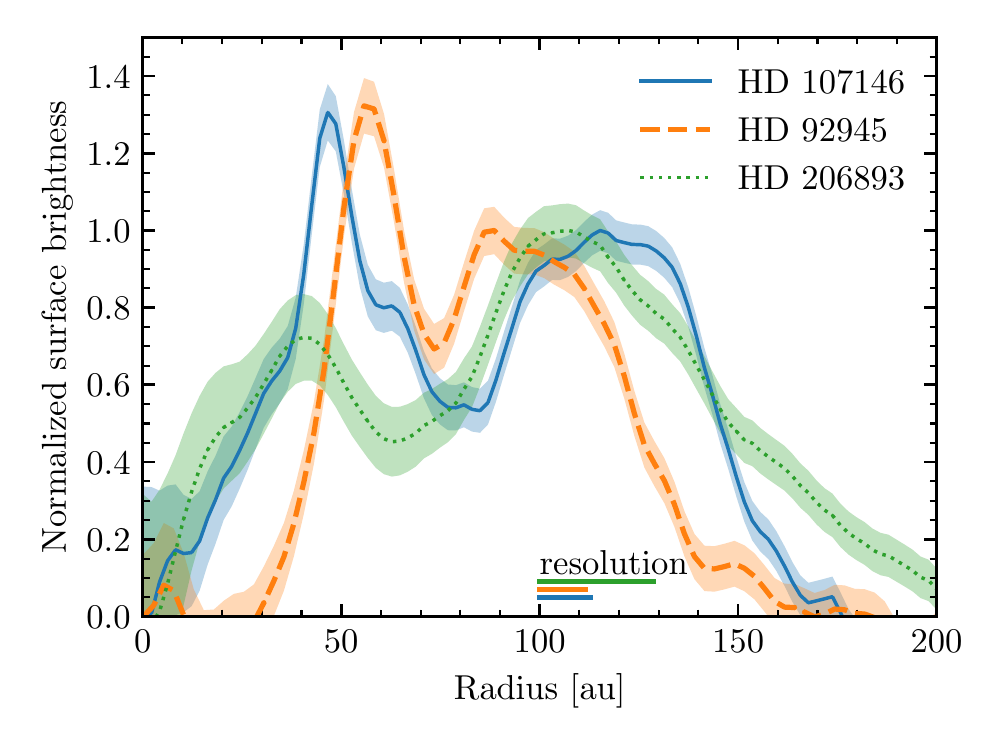}
  \caption{Deprojected surface brightness profiles of HD~107146 (blue
    line), HD~92945 (orange dashed line) and HD~206893 at 1.3 mm
    (green dotted line), computed by azimuthally averaging the
    emission. We subtracted the emission at the center using Gaussian
    profiles of FWHM equal to the beam major axes. The shaded regions
    correspond to $1\sigma$ uncertainties. Note that the shaded
    regions are representative of the uncertainty over a resolution
    element, whose size are represented by coloured horizontal lines
    at the bottom of the figure. The dashed vertical line represents
    the semi-major axis of HD~206893~B and the grey region its chaotic
    zone.}
 \label{fig:comparison}
\end{figure}

Although the gap location of all four debris discs with gaps observed
with ALMA lie at a similar radius, they have significant differences
in their radial profiles. In Figure \ref{fig:comparison} we compare
the surface brightness of these discs (except for HD15115 that is edge
on). HD~206893's disc seems to be the widest disc with the smallest
inner edge and largest outer edge. Given the lower S/N in HD~206893
observations, it is hard to compare the gap width and depth with the
other two. Nevertheless, a rough comparison indicates that HD~206893's
gap is wider than HD~92945's and deeper than HD~107146 (the only gap
that is well resolved with several beams across). Another strong
difference is that in HD~206893, the peak in surface density is beyond
the gap, while this is the opposite in HD~107146 and HD~92945. This
difference could be due to collisional evolution if HD~206893 is
significantly older (e.g. 700 Myr old) compared to HD~92945 and
HD~107146 (e.g. 100~Myr old), in which case we expect a larger
relative depletion between the region interior and exterior to the gap
\citep[e.g.][]{Wyatt2007hotdust, Kennedy2010}.

In scattered light, gaps have also been detected in the distribution
of small grains in 6 exoKuiper belts: HD~92945 and HD~15115 with gaps
also seen at mm wavelengths \citep[]{Golimowski2011, Schneider2014,
  Engler2019}, and in HD~141569, HD~131835, HD~120326 and HD~141943,
only seen so far in scattered light observations
\citep[][]{Perrot2016, Feldt2017, Bonnefoy2017, Boccaletti2019}. These
gaps would be consistent with gaps in the distribution of
planetesimals. However, these could also be due to gas-dust
interactions, which could generate multiple rings and gaps in the
distribution of $\mu$m-sized dust \citep{Lyra2013photoelectric,
  Richert2017}. In fact, two of these systems are known to have large
levels of gas \citep[HD~141569 and HD~131835,][]{Zuckerman1995,
  Moor2015gas, Kral2019}. ALMA observations constraining the presence
of gas and the existence of these gaps in the distribution of large
dust are crucial to assess if these gaps are also present in the
distribution of planetesimals.

An important consequence of gaps being common among exoKuiper belts,
is that they indicate that a large fraction of their mass was
scattered away (unless they are inherited from protoplanetary
discs). As discussed in \cite{Marino2019}, this material could
encounter additional inner planets (e.g. at the belt inner edges)
which could scatter the material even closer in until being accreted
by temperate low mass planets \citep[e.g.][]{Marino2018scat}. Such
accretion could lead to volatile delivery \citep{Kral2018trappist1,
  Wyatt2020} and the build up of secondary atmospheres, or instead
atmospheric erosion depending on planetesimal
properties. Understanding the frequency of this process is therefore
important to constrain the evolution of atmospheres of close in
planets in systems with exoKuiper belts.







\section{Conclusions}
\label{sec:conclusions}

In this work we have presented the first ALMA observations (at 0.88
and 1.3 mm) of the system HD~206893 to image its debris disc. This
system is known to host a directly imaged companion with a mass of
12--50 \Mjup\ and a semi-major axis of 11~au (HD~206893~B), and likely
an additional inner companion at around 2~au responsible for a radial
velocity trend and proper motion anomaly (HD~206893~C). Through
analysis in the image and visibility space we have found that the disc
extends from roughly 30 to 180~au, with a peak in surface brightness
and density at 110~au, and a local minimum or gap at 74 au. This gap
is found to be $27\pm5$~au wide, which if carved by a planet in situ
through scattering can be translated to a planet mass of
$0.9^{+0.8}_{-0.5}$~\Mjup\ (HD~206893~b). This gap in a
debris/planetesimal disc is the fourth to be found, and is centred at
a similar radius as the rest, namely around 70~au. Why these gaps seem
to be located at the same radius is unclear.

In addition to studying the radial structure, we searched for
asymmetries. We find a marginal evidence of an asymmetry in the disc
with the NE half being $\sim30\%$ brighter, which cannot be explained
by a single background galaxy. If real, this asymmetry could be due to
dynamical interactions with b. We also searched for CO emission in the
system, but we found no significant emission---still consistent with
solids being volatile rich and having compositions similar to Solar
System comets.

Since it has been proposed that B's spectrum is reddened by
circumstellar material, we searched for dust emission at B's
position. We did not find any emission, which rules out the presence
of a massive dusty disc larger than 0.1~au. Moreover, we find that
accreted dust from the outer debris discs would also be insufficient
to cause any reddening. Therefore if reddening is caused by dust, this
is probably lifted inside and above its photosphere.

Based on the derived disc orientation, we were able to better
constrain the orbit of B by assuming it is co-planar with the disc. We
find B is likely on an eccentric orbit with an eccentricity of
$0.14^{+0.05}_{-0.04}$ and semi-major axis of
$11.4^{+1.1}_{-0.8}$~au. Given these constraints and B's estimated
mass, the disc could have been truncated by B, explaining its observed
inner edge of $27\pm5$~au. However, the exact position of the disc
inner edge and the predicted truncation radius (given by B's mass and
orbit) are still very uncertain, hence it is still possible that the
inner edge is farther away than expected.

We have used all available dynamical and observational constraints
(RV, proper motion anomaly, stability) to determine the mass and
semi-major axis of C. We have found that its semi-major axis should be
in the range of 1.4-4.5~au (consistent with previous work) and have a
mass between 4 and 100~\Mjup. Therefore, C could be a massive gas
giant, brown dwarf, or a low mass star. Based on their estimated
orbits and masses, secular interactions between C and B could place a
secular resonance at about 30~au, near the disc inner edge. Therefore
it is plausible that the disc inner edge was truncated by this
resonance.

While the gap at 74~au could have been carved by a planet in situ,
there are other dynamical mechanisms by which a planet could carve
such a gap. Namely secular resonances with two inner planets and
secular interactions between the disc and a planet on a highly
eccentric orbit. For the former scenario to work, the putative outer
planet would need to be located at 30~au and have a similar mass
compared to B. However, such a planet is ruled-out by direct imaging
observations which did not detect additional companions beyond B with
a similar mass. The latter scenario could work if the scattered planet
has a mass similar to the disc ($\sim$20--120~\Me), which rules out
that B is responsible for the gap. If B instead scattered out a low
mass planet, such a planet could have opened the gap and today reside
at $\sim40$~au on a low eccentricity orbit.

Since the observed dust indicates an ongoing collisional cascade and
thus ongoing destructive planetesimal collisions, the disc must have
been stirred in the past. This could have either happened via secular
interactions with B or b, or via self-stirring. We find that both
mechanisms could be efficient at stirring the disc in timescales
shorter than the age of the system, but it is likely that planet
stirring by b or B dominate.

HD~206893 is a unique laboratory to study planetary dynamics and the
interaction between planetesimal discs and massive companions. Future
deeper ALMA observations could better constrain the dust distribution
within the gap, the level of asymmetry in the disc and the exact
position of the disc inner edge. Such constraints could favour a
specific scenario of the ones discussed in this paper to explain the
gap, and help to estimate better the masses of the inner
companions. Finally, there is growing evidence indicating that gaps
could be common in exoKuiper belts, although the sample size of debris
discs that have been observed with ALMA with enough resolution and
sensitivity is still very small. If gaps carved by planets in
exoKuiper belts are common, it is possible that inward scattering of
volatile-rich material from the belt to inner planets and subsequent
accretion of volatiles is a frequent process in exoplanetary systems.

\section*{Acknowledgements}

We thank Jeff Jennings and Richard Booth for their help and discussion
on how to use Frankenstein. This paper makes use of the following ALMA
data: ADS/JAO.ALMA\#2017.1.00828.S and 2017.1.00825.S. ALMA is a
partnership of ESO (representing its member states), NSF (USA) and
NINS (Japan), together with NRC (Canada), MOST and ASIAA (Taiwan), and
KASI (Republic of Korea), in cooperation with the Republic of
Chile. The Joint ALMA Observatory is operated by ESO, AUI/NRAO and
NAOJ. T.H. acknowledges support from the European Research Council
under the Horizon 2020 Framework Program via the ERC Advanced Grant
Origins 83 24 28. GMK is supported by the Royal Society as a Royal
Society University Research Fellow. VF's postdoctoral fellowship is
supported by the Exoplanet Science Initiative at the Jet Propulsion
Laboratory, California Institute of Technology, under a contract with
the National Aeronautics and Space Administration
(80NM0018D0004). Finally, we would also like to thank the anonymous
referee for a very constructive report which improved the clarity of
this paper.


\section*{Data availability}
\textit{The data underlying this article will be shared on reasonable
  request to the corresponding author. The ALMA band 6 data is
  publicly available and can be queried and downloaded directly from
  the ALMA archive at https://almascience.nrao.edu/asax/. The band 7
  data will become publicly available on 13 August 2020 at the same
  web address.}




\bibliographystyle{mnras}
\bibliography{SM_pformation} 

\newcommand{\noop}[1]{}
\begin{thebibliography}{}
\makeatletter
\relax
\def\mn@urlcharsother{\let\do\@makeother \do\$\do\&\do\#\do\^\do\_\do\%\do\~}
\def\mn@doi{\begingroup\mn@urlcharsother \@ifnextchar [ {\mn@doi@}
  {\mn@doi@[]}}
\def\mn@doi@[#1]#2{\def\@tempa{#1}\ifx\@tempa\@empty \href
  {http://dx.doi.org/#2} {doi:#2}\else \href {http://dx.doi.org/#2} {#1}\fi
  \endgroup}
\def\mn@eprint#1#2{\mn@eprint@#1:#2::\@nil}
\def\mn@eprint@arXiv#1{\href {http://arxiv.org/abs/#1} {{\tt arXiv:#1}}}
\def\mn@eprint@dblp#1{\href {http://dblp.uni-trier.de/rec/bibtex/#1.xml}
  {dblp:#1}}
\def\mn@eprint@#1:#2:#3:#4\@nil{\def\@tempa {#1}\def\@tempb {#2}\def\@tempc
  {#3}\ifx \@tempc \@empty \let \@tempc \@tempb \let \@tempb \@tempa \fi \ifx
  \@tempb \@empty \def\@tempb {arXiv}\fi \@ifundefined
  {mn@eprint@\@tempb}{\@tempb:\@tempc}{\expandafter \expandafter \csname
  mn@eprint@\@tempb\endcsname \expandafter{\@tempc}}}

\bibitem[\protect\citeauthoryear{{Acke} et~al.,}{{Acke}
  et~al.}{2012}]{Acke2012}
{Acke} B.,  et~al., 2012, \mn@doi [\aap] {10.1051/0004-6361/201118581}, \href
  {http://adsabs.harvard.edu/abs/2012A%26A...540A.125A} {540, A125}

\bibitem[\protect\citeauthoryear{{Andrews} et~al.,}{{Andrews}
  et~al.}{2018}]{Andrews2018}
{Andrews} S.~M.,  et~al., 2018, \mn@doi [\apjl] {10.3847/2041-8213/aaf741},
  \href {https://ui.adsabs.harvard.edu/abs/2018ApJ...869L..41A} {869, L41}

\bibitem[\protect\citeauthoryear{{Baraffe}, {Chabrier}, {Barman}, {Allard}  \&
  {Hauschildt}}{{Baraffe} et~al.}{2003}]{Baraffe2003}
{Baraffe} I.,  {Chabrier} G.,  {Barman} T.~S.,  {Allard} F.,   {Hauschildt}
  P.~H.,  2003, \mn@doi [\aap] {10.1051/0004-6361:20030252}, \href
  {https://ui.adsabs.harvard.edu/abs/2003A&A...402..701B} {402, 701}

\bibitem[\protect\citeauthoryear{{Beust} et~al.,}{{Beust}
  et~al.}{2014}]{Beust2014}
{Beust} H.,  et~al., 2014, \mn@doi [\aap] {10.1051/0004-6361/201322229}, \href
  {http://adsabs.harvard.edu/abs/2014A%26A...561A..43B} {561, A43}

\bibitem[\protect\citeauthoryear{{Blunt} et~al.,}{{Blunt}
  et~al.}{2020}]{Blunt2020}
{Blunt} S.,  et~al., 2020, \mn@doi [\aj] {10.3847/1538-3881/ab6663}, \href
  {https://ui.adsabs.harvard.edu/abs/2020AJ....159...89B} {159, 89}

\bibitem[\protect\citeauthoryear{{Boccaletti} et~al.,}{{Boccaletti}
  et~al.}{2019}]{Boccaletti2019}
{Boccaletti} A.,  et~al., 2019, \mn@doi [\aap] {10.1051/0004-6361/201935135},
  \href {https://ui.adsabs.harvard.edu/abs/2019A&A...625A..21B} {625, A21}

\bibitem[\protect\citeauthoryear{{Boley}, {Payne}, {Corder}, {Dent}, {Ford}  \&
  {Shabram}}{{Boley} et~al.}{2012}]{Boley2012}
{Boley} A.~C.,  {Payne} M.~J.,  {Corder} S.,  {Dent} W.~R.~F.,  {Ford} E.~B.,
  {Shabram} M.,  2012, \mn@doi [\apjl] {10.1088/2041-8205/750/1/L21}, \href
  {http://adsabs.harvard.edu/abs/2012ApJ...750L..21B} {750, L21}

\bibitem[\protect\citeauthoryear{{Bonnefoy} et~al.,}{{Bonnefoy}
  et~al.}{2017}]{Bonnefoy2017}
{Bonnefoy} M.,  et~al., 2017, \mn@doi [\aap] {10.1051/0004-6361/201628929},
  \href {https://ui.adsabs.harvard.edu/abs/2017A&A...597L...7B} {597, L7}

\bibitem[\protect\citeauthoryear{{Bonsor}, {Wyatt}, {Kral}, {Kennedy},
  {Shannon}  \& {Ertel}}{{Bonsor} et~al.}{2018}]{Bonsor2018}
{Bonsor} A.,  {Wyatt} M.~C.,  {Kral} Q.,  {Kennedy} G.,  {Shannon} A.,
  {Ertel} S.,  2018, \mn@doi [\mnras] {10.1093/mnras/sty2200}, \href
  {https://ui.adsabs.harvard.edu/abs/2018MNRAS.480.5560B} {480, 5560}

\bibitem[\protect\citeauthoryear{{Booth}, {Wyatt}, {Morbidelli},
  {Moro-Mart{\'{\i}}n}  \& {Levison}}{{Booth} et~al.}{2009}]{Booth2009}
{Booth} M.,  {Wyatt} M.~C.,  {Morbidelli} A.,  {Moro-Mart{\'{\i}}n} A.,
  {Levison} H.~F.,  2009, \mn@doi [\mnras] {10.1111/j.1365-2966.2009.15286.x},
  \href {http://adsabs.harvard.edu/abs/2009MNRAS.399..385B} {399, 385}

\bibitem[\protect\citeauthoryear{{Booth} et~al.,}{{Booth}
  et~al.}{2016}]{Booth2016}
{Booth} M.,  et~al., 2016, \mn@doi [\mnras] {10.1093/mnrasl/slw040}, \href
  {http://adsabs.harvard.edu/abs/2016MNRAS.460L..10B} {460, L10}

\bibitem[\protect\citeauthoryear{{Boss}}{{Boss}}{1997}]{Boss1997}
{Boss} A.~P.,  1997, \mn@doi [Science] {10.1126/science.276.5320.1836}, \href
  {http://adsabs.harvard.edu/abs/1997Sci...276.1836B} {276, 1836}

\bibitem[\protect\citeauthoryear{{Boss}}{{Boss}}{2003}]{Boss2003}
{Boss} A.~P.,  2003, \mn@doi [\apj] {10.1086/379163}, \href
  {https://ui.adsabs.harvard.edu/abs/2003ApJ...599..577B} {599, 577}

\bibitem[\protect\citeauthoryear{{Boss}}{{Boss}}{2011}]{Boss2011}
{Boss} A.~P.,  2011, \mn@doi [\apj] {10.1088/0004-637X/731/1/74}, \href
  {https://ui.adsabs.harvard.edu/abs/2011ApJ...731...74B} {731, 74}

\bibitem[\protect\citeauthoryear{{Burns}, {Lamy}  \& {Soter}}{{Burns}
  et~al.}{1979}]{Burns1979}
{Burns} J.~A.,  {Lamy} P.~L.,   {Soter} S.,  1979, \mn@doi [\icarus]
  {10.1016/0019-1035(79)90050-2}, \href
  {http://adsabs.harvard.edu/abs/1979Icar...40....1B} {40, 1}

\bibitem[\protect\citeauthoryear{{Chen}, {Mittal}, {Kuchner}, {Forrest},
  {Lisse}, {Manoj}, {Sargent}  \& {Watson}}{{Chen} et~al.}{2014}]{Chen2014}
{Chen} C.~H.,  {Mittal} T.,  {Kuchner} M.,  {Forrest} W.~J.,  {Lisse} C.~M.,
  {Manoj} P.,  {Sargent} B.~A.,   {Watson} D.~M.,  2014, \mn@doi [\apjs]
  {10.1088/0067-0049/211/2/25}, \href
  {http://adsabs.harvard.edu/abs/2014ApJS..211...25C} {211, 25}

\bibitem[\protect\citeauthoryear{{Chiang}, {Kite}, {Kalas}, {Graham}  \&
  {Clampin}}{{Chiang} et~al.}{2009}]{Chiang2009}
{Chiang} E.,  {Kite} E.,  {Kalas} P.,  {Graham} J.~R.,   {Clampin} M.,  2009,
  \mn@doi [\apj] {10.1088/0004-637X/693/1/734}, \href
  {https://ui.adsabs.harvard.edu/abs/2009ApJ...693..734C} {693, 734}

\bibitem[\protect\citeauthoryear{{Daley} et~al.,}{{Daley}
  et~al.}{2019}]{Daley2019}
{Daley} C.,  et~al., 2019, \mn@doi [\apj] {10.3847/1538-4357/ab1074}, \href
  {https://ui.adsabs.harvard.edu/abs/2019ApJ...875...87D} {875, 87}

\bibitem[\protect\citeauthoryear{{Delorme} et~al.,}{{Delorme}
  et~al.}{2017}]{Delorme2017}
{Delorme} P.,  et~al., 2017, \mn@doi [\aap] {10.1051/0004-6361/201731145},
  \href {https://ui.adsabs.harvard.edu/abs/2017A&A...608A..79D} {608, A79}

\bibitem[\protect\citeauthoryear{{Dent} et~al.,}{{Dent}
  et~al.}{2014}]{Dent2014}
{Dent} W.~R.~F.,  et~al., 2014, \mn@doi [Science] {10.1126/science.1248726},
  \href {http://adsabs.harvard.edu/abs/2014Sci...343.1490D} {343, 1490}

\bibitem[\protect\citeauthoryear{{Dullemond} \& {Penzlin}}{{Dullemond} \&
  {Penzlin}}{2018}]{Dullemond2018}
{Dullemond} C.~P.,  {Penzlin} A.~B.~T.,  2018, \mn@doi [\aap]
  {10.1051/0004-6361/201731878}, \href
  {http://adsabs.harvard.edu/abs/2018A%26A...609A..50D} {609, A50}

\bibitem[\protect\citeauthoryear{{Dullemond}, {Juhasz}, {Pohl}, {Sereshti},
  {Shetty}, {Peters}, {Commercon}  \& {Flock}}{{Dullemond}
  et~al.}{2017}]{radmc3d}
{Dullemond} C.,  {Juhasz} A.,  {Pohl} A.,  {Sereshti} F.,  {Shetty} R.,
  {Peters} T.,  {Commercon} B.,   {Flock} M.,  2017, RADMC3D v0.41
  {http://www.ita.uni-heidelberg.de/ dullemond/software/radmc-3d/}

\bibitem[\protect\citeauthoryear{{Engler} et~al.,}{{Engler}
  et~al.}{2019}]{Engler2019}
{Engler} N.,  et~al., 2019, \mn@doi [\aap] {10.1051/0004-6361/201833542}, \href
  {https://ui.adsabs.harvard.edu/abs/2019A&A...622A.192E} {622, A192}

\bibitem[\protect\citeauthoryear{{Faramaz}, {Beust}, {Augereau}, {Kalas}  \&
  {Graham}}{{Faramaz} et~al.}{2015}]{Faramaz2015}
{Faramaz} V.,  {Beust} H.,  {Augereau} J.-C.,  {Kalas} P.,   {Graham} J.~R.,
  2015, \mn@doi [\aap] {10.1051/0004-6361/201424691}, \href
  {http://adsabs.harvard.edu/abs/2015A%26A...573A..87F} {573, A87}

\bibitem[\protect\citeauthoryear{{Faramaz} et~al.,}{{Faramaz}
  et~al.}{2019}]{Faramaz2019}
{Faramaz} V.,  et~al., 2019, \mn@doi [\aj] {10.3847/1538-3881/ab3ec1}, \href
  {https://ui.adsabs.harvard.edu/abs/2019AJ....158..162F} {158, 162}

\bibitem[\protect\citeauthoryear{{Feldt} et~al.,}{{Feldt}
  et~al.}{2017}]{Feldt2017}
{Feldt} M.,  et~al., 2017, \mn@doi [\aap] {10.1051/0004-6361/201629261}, \href
  {http://adsabs.harvard.edu/abs/2017A%26A...601A...7F} {601, A7}

\bibitem[\protect\citeauthoryear{{Flock}, {Ruge}, {Dzyurkevich}, {Henning},
  {Klahr}  \& {Wolf}}{{Flock} et~al.}{2015}]{Flock2015}
{Flock} M.,  {Ruge} J.~P.,  {Dzyurkevich} N.,  {Henning} T.,  {Klahr} H.,
  {Wolf} S.,  2015, \mn@doi [\aap] {10.1051/0004-6361/201424693}, \href
  {https://ui.adsabs.harvard.edu/#abs/2015A&A...574A..68F} {574, A68}

\bibitem[\protect\citeauthoryear{{Foreman-Mackey}}{{Foreman-Mackey}}{2016}]{corner}
{Foreman-Mackey} D.,  2016, \mn@doi [The Journal of Open Source Software]
  {10.21105/joss.00024}, \href
  {https://ui.adsabs.harvard.edu/abs/2016JOSS....1...24F} {1, 24}

\bibitem[\protect\citeauthoryear{{Fujimoto}, {Ouchi}, {Shibuya}  \&
  {Nagai}}{{Fujimoto} et~al.}{2017}]{Fujimoto2017}
{Fujimoto} S.,  {Ouchi} M.,  {Shibuya} T.,   {Nagai} H.,  2017, \mn@doi [\apj]
  {10.3847/1538-4357/aa93e6}, \href
  {https://ui.adsabs.harvard.edu/abs/2017ApJ...850...83F} {850, 83}

\bibitem[\protect\citeauthoryear{{Gaia Collaboration} et~al.,}{{Gaia
  Collaboration} et~al.}{2018}]{Gaiadr2}
{Gaia Collaboration} et~al., 2018, \mn@doi [\aap]
  {10.1051/0004-6361/201833051}, \href
  {http://adsabs.harvard.edu/abs/2018A%26A...616A...1G} {616, A1}

\bibitem[\protect\citeauthoryear{{Gaspar} \& {Rieke}}{{Gaspar} \&
  {Rieke}}{2020}]{Gaspar2020}
{Gaspar} A.,  {Rieke} G.~H.,  2020, arXiv e-prints, \href
  {https://ui.adsabs.harvard.edu/abs/2020arXiv200408736G} {p. arXiv:2004.08736}

\bibitem[\protect\citeauthoryear{{Geiler}, {Krivov}, {Booth}  \&
  {L{\"o}hne}}{{Geiler} et~al.}{2019}]{Geiler2019}
{Geiler} F.,  {Krivov} A.~V.,  {Booth} M.,   {L{\"o}hne} T.,  2019, \mn@doi
  [\mnras] {10.1093/mnras/sty3160}, \href
  {https://ui.adsabs.harvard.edu/abs/2019MNRAS.483..332G} {483, 332}

\bibitem[\protect\citeauthoryear{{Gladman}}{{Gladman}}{1993}]{Gladman1993}
{Gladman} B.,  1993, \mn@doi [\icarus] {10.1006/icar.1993.1169}, \href
  {https://ui.adsabs.harvard.edu/abs/1993Icar..106..247G} {106, 247}

\bibitem[\protect\citeauthoryear{{Golimowski} et~al.,}{{Golimowski}
  et~al.}{2011}]{Golimowski2011}
{Golimowski} D.~A.,  et~al., 2011, \mn@doi [\aj] {10.1088/0004-6256/142/1/30},
  \href {http://adsabs.harvard.edu/abs/2011AJ....142...30G} {142, 30}

\bibitem[\protect\citeauthoryear{{Grandjean} et~al.,}{{Grandjean}
  et~al.}{2019}]{Grandjean2019}
{Grandjean} A.,  et~al., 2019, \mn@doi [\aap] {10.1051/0004-6361/201935044},
  \href {https://ui.adsabs.harvard.edu/abs/2019A&A...627L...9G} {627, L9}

\bibitem[\protect\citeauthoryear{{Greaves}, {Holland}, {Jayawardhana}, {Wyatt}
  \& {Dent}}{{Greaves} et~al.}{2004}]{Greaves2004}
{Greaves} J.~S.,  {Holland} W.~S.,  {Jayawardhana} R.,  {Wyatt} M.~C.,   {Dent}
  W.~R.~F.,  2004, \mn@doi [\mnras] {10.1111/j.1365-2966.2004.07440.x}, \href
  {http://adsabs.harvard.edu/abs/2004MNRAS.348.1097G} {348, 1097}

\bibitem[\protect\citeauthoryear{{Huang} et~al.,}{{Huang}
  et~al.}{2018}]{Huang2018}
{Huang} J.,  et~al., 2018, \mn@doi [\apjl] {10.3847/2041-8213/aaf740}, \href
  {https://ui.adsabs.harvard.edu/abs/2018ApJ...869L..42H} {869, L42}

\bibitem[\protect\citeauthoryear{{Hughes}, {Duch{\^e}ne}  \&
  {Matthews}}{{Hughes} et~al.}{2018}]{Hughes2018}
{Hughes} A.~M.,  {Duch{\^e}ne} G.,   {Matthews} B.~C.,  2018, \mn@doi [\araa]
  {10.1146/annurev-astro-081817-052035}, \href
  {http://adsabs.harvard.edu/abs/2018ARA%26A..56..541H} {56, 541}

\bibitem[\protect\citeauthoryear{{Jennings}, {Booth}, {Tazzari}, {Rosotti}  \&
  {Clarke}}{{Jennings} et~al.}{2020}]{Jennings2020}
{Jennings} J.,  {Booth} R.~A.,  {Tazzari} M.,  {Rosotti} G.~P.,   {Clarke}
  C.~J.,  2020, \mn@doi [\mnras] {10.1093/mnras/staa1365}, \href
  {https://ui.adsabs.harvard.edu/abs/2020MNRAS.tmp.1491J} {}

\bibitem[\protect\citeauthoryear{{Kalas}, {Graham}  \& {Clampin}}{{Kalas}
  et~al.}{2005}]{Kalas2005}
{Kalas} P.,  {Graham} J.~R.,   {Clampin} M.,  2005, \mn@doi [\nat]
  {10.1038/nature03601}, \href
  {http://adsabs.harvard.edu/abs/2005Natur.435.1067K} {435, 1067}

\bibitem[\protect\citeauthoryear{{Kalas} et~al.,}{{Kalas}
  et~al.}{2008}]{Kalas2008}
{Kalas} P.,  et~al., 2008, \mn@doi [Science] {10.1126/science.1166609}, \href
  {http://adsabs.harvard.edu/abs/2008Sci...322.1345K} {322, 1345}

\bibitem[\protect\citeauthoryear{{Kennedy} \& {Wyatt}}{{Kennedy} \&
  {Wyatt}}{2010}]{Kennedy2010}
{Kennedy} G.~M.,  {Wyatt} M.~C.,  2010, \mn@doi [\mnras]
  {10.1111/j.1365-2966.2010.16528.x}, \href
  {http://adsabs.harvard.edu/abs/2010MNRAS.405.1253K} {405, 1253}

\bibitem[\protect\citeauthoryear{{Kennedy} \& {Wyatt}}{{Kennedy} \&
  {Wyatt}}{2011}]{Kennedy2011}
{Kennedy} G.~M.,  {Wyatt} M.~C.,  2011, \mn@doi [\mnras]
  {10.1111/j.1365-2966.2010.18041.x}, \href
  {http://adsabs.harvard.edu/abs/2011MNRAS.412.2137K} {412, 2137}

\bibitem[\protect\citeauthoryear{{Kennedy} et~al.,}{{Kennedy}
  et~al.}{2015a}]{Kennedy2015lbti}
{Kennedy} G.~M.,  et~al., 2015a, \mn@doi [\apjs] {10.1088/0067-0049/216/2/23},
  \href {http://adsabs.harvard.edu/abs/2015ApJS..216...23K} {216, 23}

\bibitem[\protect\citeauthoryear{{Kennedy} et~al.,}{{Kennedy}
  et~al.}{2015b}]{Kennedy2015superearths}
{Kennedy} G.~M.,  et~al., 2015b, \mn@doi [\mnras] {10.1093/mnras/stv511}, \href
  {http://adsabs.harvard.edu/abs/2015MNRAS.449.3121K} {449, 3121}

\bibitem[\protect\citeauthoryear{{Kenyon} \& {Bromley}}{{Kenyon} \&
  {Bromley}}{2008}]{Kenyon2008}
{Kenyon} S.~J.,  {Bromley} B.~C.,  2008, \mn@doi [\apjs] {10.1086/591794},
  \href {http://adsabs.harvard.edu/abs/2008ApJS..179..451K} {179, 451}

\bibitem[\protect\citeauthoryear{{Kenyon} \& {Bromley}}{{Kenyon} \&
  {Bromley}}{2010}]{Kenyon2010}
{Kenyon} S.~J.,  {Bromley} B.~C.,  2010, \mn@doi [\apjs]
  {10.1088/0067-0049/188/1/242}, \href
  {http://adsabs.harvard.edu/abs/2010ApJS..188..242K} {188, 242}

\bibitem[\protect\citeauthoryear{{Kervella}, {Arenou}, {Mignard}  \&
  {Th{\'e}venin}}{{Kervella} et~al.}{2019}]{Kervella2019}
{Kervella} P.,  {Arenou} F.,  {Mignard} F.,   {Th{\'e}venin} F.,  2019, \mn@doi
  [\aap] {10.1051/0004-6361/201834371}, \href
  {https://ui.adsabs.harvard.edu/abs/2019A&A...623A..72K} {623, A72}

\bibitem[\protect\citeauthoryear{{Klahr} \& {Lin}}{{Klahr} \&
  {Lin}}{2005}]{Klahr2005}
{Klahr} H.,  {Lin} D.~N.~C.,  2005, \mn@doi [\apj] {10.1086/432965}, \href
  {http://adsabs.harvard.edu/abs/2005ApJ...632.1113K} {632, 1113}

\bibitem[\protect\citeauthoryear{{Konopacky} et~al.,}{{Konopacky}
  et~al.}{2016}]{Konopacky2016}
{Konopacky} Q.~M.,  et~al., 2016, \mn@doi [\apjl] {10.3847/2041-8205/829/1/L4},
  \href {https://ui.adsabs.harvard.edu/abs/2016ApJ...829L...4K} {829, L4}

\bibitem[\protect\citeauthoryear{{Kral}, {Matr{\`a}}, {Wyatt}  \&
  {Kennedy}}{{Kral} et~al.}{2017}]{Kral2017CO}
{Kral} Q.,  {Matr{\`a}} L.,  {Wyatt} M.~C.,   {Kennedy} G.~M.,  2017, \mn@doi
  [\mnras] {10.1093/mnras/stx730}, \href
  {http://adsabs.harvard.edu/abs/2017MNRAS.469..521K} {469, 521}

\bibitem[\protect\citeauthoryear{{Kral}, {Wyatt}, {Triaud}, {Marino},
  {Thebault}  \& {Shorttle}}{{Kral} et~al.}{2018}]{Kral2018trappist1}
{Kral} Q.,  {Wyatt} M.~C.,  {Triaud} A.~H.~M.~J.,  {Marino} S.,  {Thebault} P.,
    {Shorttle} O.,  2018, preprint, \href
  {http://adsabs.harvard.edu/abs/2018arXiv180205034K} {} (\mn@eprint {arXiv}
  {1802.05034})

\bibitem[\protect\citeauthoryear{{Kral}, {Marino}, {Wyatt}, {Kama}  \&
  {Matr{\`a}}}{{Kral} et~al.}{2019}]{Kral2019}
{Kral} Q.,  {Marino} S.,  {Wyatt} M.~C.,  {Kama} M.,   {Matr{\`a}} L.,  2019,
  \mn@doi [\mnras] {10.1093/mnras/sty2923}, \href
  {https://ui.adsabs.harvard.edu/abs/2019MNRAS.489.3670K} {489, 3670}

\bibitem[\protect\citeauthoryear{{Krivov} \& {Booth}}{{Krivov} \&
  {Booth}}{2018}]{Krivov2018stirring}
{Krivov} A.~V.,  {Booth} M.,  2018, \mn@doi [\mnras] {10.1093/mnras/sty1607},
  \href {https://ui.adsabs.harvard.edu/abs/2018MNRAS.479.3300K} {479, 3300}

\bibitem[\protect\citeauthoryear{{Lagrange} et~al.,}{{Lagrange}
  et~al.}{2012}]{Lagrange2012}
{Lagrange} A.~M.,  et~al., 2012, \mn@doi [\aap] {10.1051/0004-6361/201118274},
  \href {https://ui.adsabs.harvard.edu/abs/2012A&A...542A..40L} {542, A40}

\bibitem[\protect\citeauthoryear{{Lagrange} et~al.,}{{Lagrange}
  et~al.}{2019}]{Lagrange2019}
{Lagrange} A.~M.,  et~al., 2019, \mn@doi [\aap] {10.1051/0004-6361/201834302},
  \href {https://ui.adsabs.harvard.edu/abs/2019A&A...621L...8L} {621, L8}

\bibitem[\protect\citeauthoryear{{Launhardt} et~al.,}{{Launhardt}
  et~al.}{2020}]{Launhardt2020}
{Launhardt} R.,  et~al., 2020, \mn@doi [\aap] {10.1051/0004-6361/201937000},
  \href {https://ui.adsabs.harvard.edu/abs/2020A&A...635A.162L} {635, A162}

\bibitem[\protect\citeauthoryear{{Lazzoni} et~al.,}{{Lazzoni}
  et~al.}{2018}]{Lazzoni2018}
{Lazzoni} C.,  et~al., 2018, \mn@doi [\aap] {10.1051/0004-6361/201731426},
  \href {http://adsabs.harvard.edu/abs/2018A%26A...611A..43L} {611, A43}

\bibitem[\protect\citeauthoryear{{Lindroos} et~al.,}{{Lindroos}
  et~al.}{2016}]{Lindroos2016}
{Lindroos} L.,  et~al., 2016, \mn@doi [\mnras] {10.1093/mnras/stw1628}, \href
  {https://ui.adsabs.harvard.edu/abs/2016MNRAS.462.1192L} {462, 1192}

\bibitem[\protect\citeauthoryear{{Long} et~al.,}{{Long}
  et~al.}{2018}]{Long2018}
{Long} F.,  et~al., 2018, \mn@doi [\apj] {10.3847/1538-4357/aae8e1}, \href
  {https://ui.adsabs.harvard.edu/abs/2018ApJ...869...17L} {869, 17}

\bibitem[\protect\citeauthoryear{{Long} et~al.,}{{Long}
  et~al.}{2019}]{Long2019}
{Long} F.,  et~al., 2019, \mn@doi [\apj] {10.3847/1538-4357/ab2d2d}, \href
  {https://ui.adsabs.harvard.edu/abs/2019ApJ...882...49L} {882, 49}

\bibitem[\protect\citeauthoryear{{Lor{\'e}n-Aguilar} \&
  {Bate}}{{Lor{\'e}n-Aguilar} \& {Bate}}{2015}]{Loren-Aguilar2015}
{Lor{\'e}n-Aguilar} P.,  {Bate} M.~R.,  2015, \mn@doi [\mnras]
  {10.1093/mnrasl/slv109}, \href
  {https://ui.adsabs.harvard.edu/abs/2015MNRAS.453L..78L} {453, L78}

\bibitem[\protect\citeauthoryear{{Lyra} \& {Kuchner}}{{Lyra} \&
  {Kuchner}}{2013}]{Lyra2013photoelectric}
{Lyra} W.,  {Kuchner} M.,  2013, \mn@doi [\nat] {10.1038/nature12281}, \href
  {http://adsabs.harvard.edu/abs/2013Natur.499..184L} {499, 184}

\bibitem[\protect\citeauthoryear{{MacGregor} et~al.,}{{MacGregor}
  et~al.}{2016}]{MacGregor2016}
{MacGregor} M.~A.,  et~al., 2016, \mn@doi [\apj] {10.3847/0004-637X/823/2/79},
  \href {https://ui.adsabs.harvard.edu/abs/2016ApJ...823...79M} {823, 79}

\bibitem[\protect\citeauthoryear{{MacGregor} et~al.,}{{MacGregor}
  et~al.}{2017}]{MacGregor2017}
{MacGregor} M.~A.,  et~al., 2017, \mn@doi [\apj] {10.3847/1538-4357/aa71ae},
  \href {http://adsabs.harvard.edu/abs/2017ApJ...842....8M} {842, 8}

\bibitem[\protect\citeauthoryear{{MacGregor} et~al.,}{{MacGregor}
  et~al.}{2019}]{MacGregor2019}
{MacGregor} M.~A.,  et~al., 2019, \mn@doi [\apjl] {10.3847/2041-8213/ab21c2},
  \href {https://ui.adsabs.harvard.edu/abs/2019ApJ...877L..32M} {877, L32}

\bibitem[\protect\citeauthoryear{{Marino} et~al.,}{{Marino}
  et~al.}{2016}]{Marino2016}
{Marino} S.,  et~al., 2016, \mn@doi [\mnras] {10.1093/mnras/stw1216}, \href
  {http://adsabs.harvard.edu/abs/2016MNRAS.460.2933M} {460, 2933}

\bibitem[\protect\citeauthoryear{{Marino} et~al.,}{{Marino}
  et~al.}{2017a}]{Marino2017etacorvi}
{Marino} S.,  et~al., 2017a, \mn@doi [\mnras] {10.1093/mnras/stw2867}, \href
  {http://adsabs.harvard.edu/abs/2017MNRAS.465.2595M} {465, 2595}

\bibitem[\protect\citeauthoryear{{Marino}, {Wyatt}, {Kennedy}, {Holland},
  {Matr{\`a}}, {Shannon}  \& {Ivison}}{{Marino}
  et~al.}{2017b}]{Marino201761vir}
{Marino} S.,  {Wyatt} M.~C.,  {Kennedy} G.~M.,  {Holland} W.,  {Matr{\`a}} L.,
  {Shannon} A.,   {Ivison} R.~J.,  2017b, \mn@doi [\mnras]
  {10.1093/mnras/stx1102}, \href
  {http://adsabs.harvard.edu/abs/2017MNRAS.469.3518M} {469, 3518}

\bibitem[\protect\citeauthoryear{{Marino} et~al.,}{{Marino}
  et~al.}{2018a}]{Marino2018hd107}
{Marino} S.,  et~al., 2018a, \mn@doi [\mnras] {10.1093/mnras/sty1790}, \href
  {http://adsabs.harvard.edu/abs/2018MNRAS.479.5423M} {479, 5423}

\bibitem[\protect\citeauthoryear{{Marino}, {Bonsor}, {Wyatt}  \&
  {Kral}}{{Marino} et~al.}{2018b}]{Marino2018scat}
{Marino} S.,  {Bonsor} A.,  {Wyatt} M.~C.,   {Kral} Q.,  2018b, \mn@doi
  [\mnras] {10.1093/mnras/sty1475}, \href
  {https://ui.adsabs.harvard.edu/abs/2018MNRAS.479.1651M} {479, 1651}

\bibitem[\protect\citeauthoryear{{Marino}, {Yelverton}, {Booth}, {Faramaz},
  {Kennedy}, {Matr{\`a}}  \& {Wyatt}}{{Marino} et~al.}{2019}]{Marino2019}
{Marino} S.,  {Yelverton} B.,  {Booth} M.,  {Faramaz} V.,  {Kennedy} G.~M.,
  {Matr{\`a}} L.,   {Wyatt} M.~C.,  2019, \mn@doi [\mnras]
  {10.1093/mnras/stz049}, \href
  {https://ui.adsabs.harvard.edu/abs/2019MNRAS.484.1257M} {484, 1257}

\bibitem[\protect\citeauthoryear{{Marino}, {Flock}, {Henning}, {Kral},
  {Matr{\`a}}  \& {Wyatt}}{{Marino} et~al.}{2020}]{Marino2020gas}
{Marino} S.,  {Flock} M.,  {Henning} T.,  {Kral} Q.,  {Matr{\`a}} L.,   {Wyatt}
  M.~C.,  2020, \mn@doi [\mnras] {10.1093/mnras/stz3487}, \href
  {https://ui.adsabs.harvard.edu/abs/2020MNRAS.492.4409M} {492, 4409}

\bibitem[\protect\citeauthoryear{{Marois}, {Zuckerman}, {Konopacky},
  {Macintosh}  \& {Barman}}{{Marois} et~al.}{2010}]{Marois2010}
{Marois} C.,  {Zuckerman} B.,  {Konopacky} Q.~M.,  {Macintosh} B.,   {Barman}
  T.,  2010, \mn@doi [\nat] {10.1038/nature09684}, \href
  {http://adsabs.harvard.edu/abs/2010Natur.468.1080M} {468, 1080}

\bibitem[\protect\citeauthoryear{{Marshall} et~al.,}{{Marshall}
  et~al.}{2014}]{Marshall2014}
{Marshall} J.~P.,  et~al., 2014, \mn@doi [\aap] {10.1051/0004-6361/201323058},
  \href {https://ui.adsabs.harvard.edu/abs/2014A&A...565A..15M} {565, A15}

\bibitem[\protect\citeauthoryear{{Matr{\`a}}, {Pani{\'c}}, {Wyatt}  \&
  {Dent}}{{Matr{\`a}} et~al.}{2015}]{Matra2015}
{Matr{\`a}} L.,  {Pani{\'c}} O.,  {Wyatt} M.~C.,   {Dent} W.~R.~F.,  2015,
  \mn@doi [\mnras] {10.1093/mnras/stu2619}, \href
  {http://adsabs.harvard.edu/abs/2015MNRAS.447.3936M} {447, 3936}

\bibitem[\protect\citeauthoryear{{Matr{\`a}} et~al.,}{{Matr{\`a}}
  et~al.}{2017}]{Matra2017fomalhaut}
{Matr{\`a}} L.,  et~al., 2017, \mn@doi [\apj] {10.3847/1538-4357/aa71b4}, \href
  {http://adsabs.harvard.edu/abs/2017ApJ...842....9M} {842, 9}

\bibitem[\protect\citeauthoryear{{Matr{\`a}}, {Wilner}, {{\"O}berg}, {Andrews},
  {Loomis}, {Wyatt}  \& {Dent}}{{Matr{\`a}} et~al.}{2018a}]{Matra2018}
{Matr{\`a}} L.,  {Wilner} D.~J.,  {{\"O}berg} K.~I.,  {Andrews} S.~M.,
  {Loomis} R.~A.,  {Wyatt} M.~C.,   {Dent} W.~R.~F.,  2018a, \mn@doi [\apj]
  {10.3847/1538-4357/aaa42a}, \href
  {http://adsabs.harvard.edu/abs/2018ApJ...853..147M} {853, 147}

\bibitem[\protect\citeauthoryear{{Matr{\`a}}, {Marino}, {Kennedy}, {Wyatt},
  {{\"O}berg}  \& {Wilner}}{{Matr{\`a}} et~al.}{2018b}]{Matra2018mmlaw}
{Matr{\`a}} L.,  {Marino} S.,  {Kennedy} G.~M.,  {Wyatt} M.~C.,  {{\"O}berg}
  K.~I.,   {Wilner} D.~J.,  2018b, \mn@doi [\apj] {10.3847/1538-4357/aabcc4},
  \href {http://adsabs.harvard.edu/abs/2018ApJ...859...72M} {859, 72}

\bibitem[\protect\citeauthoryear{{Matr{\`a}}, {Wyatt}, {Wilner}, {Dent},
  {Marino}, {Kennedy}  \& {Milli}}{{Matr{\`a}} et~al.}{2019}]{Matra2019betapic}
{Matr{\`a}} L.,  {Wyatt} M.~C.,  {Wilner} D.~J.,  {Dent} W.~R.~F.,  {Marino}
  S.,  {Kennedy} G.~M.,   {Milli} J.,  2019, \mn@doi [\aj]
  {10.3847/1538-3881/ab06c0}, \href
  {https://ui.adsabs.harvard.edu/abs/2019AJ....157..135M} {157, 135}

\bibitem[\protect\citeauthoryear{{McMullin}, {Waters}, {Schiebel}, {Young}  \&
  {Golap}}{{McMullin} et~al.}{2007}]{casa}
{McMullin} J.~P.,  {Waters} B.,  {Schiebel} D.,  {Young} W.,   {Golap} K.,
  2007, in {Shaw} R.~A.,  {Hill} F.,   {Bell} D.~J.,  eds,  Astronomical
  Society of the Pacific Conference Series Vol. 376, Astronomical Data Analysis
  Software and Systems XVI. p.~127

\bibitem[\protect\citeauthoryear{{Meshkat} et~al.,}{{Meshkat}
  et~al.}{2017}]{Meshkat2017}
{Meshkat} T.,  et~al., 2017, \mn@doi [\aj] {10.3847/1538-3881/aa8e9a}, \href
  {http://adsabs.harvard.edu/abs/2017AJ....154..245M} {154, 245}

\bibitem[\protect\citeauthoryear{{Milli} et~al.,}{{Milli}
  et~al.}{2017}]{Milli2017hd206}
{Milli} J.,  et~al., 2017, \mn@doi [\aap] {10.1051/0004-6361/201629908}, \href
  {https://ui.adsabs.harvard.edu/abs/2017A&A...597L...2M} {597, L2}

\bibitem[\protect\citeauthoryear{{Mo{\'o}r}, {{\'A}brah{\'a}m}, {Derekas},
  {Kiss}, {Kiss}, {Apai}, {Grady}  \& {Henning}}{{Mo{\'o}r}
  et~al.}{2006}]{Moor2006}
{Mo{\'o}r} A.,  {{\'A}brah{\'a}m} P.,  {Derekas} A.,  {Kiss} C.,  {Kiss} L.~L.,
   {Apai} D.,  {Grady} C.,   {Henning} T.,  2006, \mn@doi [\apj]
  {10.1086/503381}, \href
  {https://ui.adsabs.harvard.edu/abs/2006ApJ...644..525M} {644, 525}

\bibitem[\protect\citeauthoryear{{Mo{\'o}r} et~al.,}{{Mo{\'o}r}
  et~al.}{2015}]{Moor2015gas}
{Mo{\'o}r} A.,  et~al., 2015, \mn@doi [\apj] {10.1088/0004-637X/814/1/42},
  \href {http://adsabs.harvard.edu/abs/2015ApJ...814...42M} {814, 42}

\bibitem[\protect\citeauthoryear{{Mo{\'o}r} et~al.,}{{Mo{\'o}r}
  et~al.}{2017}]{Moor2017}
{Mo{\'o}r} A.,  et~al., 2017, \mn@doi [\apj] {10.3847/1538-4357/aa8e4e}, \href
  {http://adsabs.harvard.edu/abs/2017ApJ...849..123M} {849, 123}

\bibitem[\protect\citeauthoryear{{Moro-Mart{\'{\i}}n}
  et~al.,}{{Moro-Mart{\'{\i}}n} et~al.}{2007}]{Moro-Martin2007}
{Moro-Mart{\'{\i}}n} A.,  et~al., 2007, \mn@doi [\apj] {10.1086/511746}, \href
  {http://adsabs.harvard.edu/abs/2007ApJ...658.1312M} {658, 1312}

\bibitem[\protect\citeauthoryear{{Moro-Mart{\'\i}n}, {Malhotra}, {Bryden},
  {Rieke}, {Su}, {Beichman}  \& {Lawler}}{{Moro-Mart{\'\i}n}
  et~al.}{2010}]{Moro-Martin2010}
{Moro-Mart{\'\i}n} A.,  {Malhotra} R.,  {Bryden} G.,  {Rieke} G.~H.,  {Su} K.
  Y.~L.,  {Beichman} C.~A.,   {Lawler} S.~M.,  2010, \mn@doi [\apj]
  {10.1088/0004-637X/717/2/1123}, \href
  {https://ui.adsabs.harvard.edu/abs/2010ApJ...717.1123M} {717, 1123}

\bibitem[\protect\citeauthoryear{{Moro-Mart{\'{\i}}n}
  et~al.,}{{Moro-Mart{\'{\i}}n} et~al.}{2015}]{Moro-Martin2015}
{Moro-Mart{\'{\i}}n} A.,  et~al., 2015, \mn@doi [\apj]
  {10.1088/0004-637X/801/2/143}, \href
  {http://adsabs.harvard.edu/abs/2015ApJ...801..143M} {801, 143}

\bibitem[\protect\citeauthoryear{{Morrison} \& {Malhotra}}{{Morrison} \&
  {Malhotra}}{2015}]{Morrison2015}
{Morrison} S.,  {Malhotra} R.,  2015, \mn@doi [\apj]
  {10.1088/0004-637X/799/1/41}, \href
  {http://adsabs.harvard.edu/abs/2015ApJ...799...41M} {799, 41}

\bibitem[\protect\citeauthoryear{{Mouillet}, {Larwood}, {Papaloizou}  \&
  {Lagrange}}{{Mouillet} et~al.}{1997}]{Mouillet1997}
{Mouillet} D.,  {Larwood} J.~D.,  {Papaloizou} J.~C.~B.,   {Lagrange} A.~M.,
  1997, \mn@doi [\mnras] {10.1093/mnras/292.4.896}, \href
  {http://adsabs.harvard.edu/abs/1997MNRAS.292..896M} {292, 896}

\bibitem[\protect\citeauthoryear{{Mumma} \& {Charnley}}{{Mumma} \&
  {Charnley}}{2011}]{Mumma2011}
{Mumma} M.~J.,  {Charnley} S.~B.,  2011, \mn@doi [\araa]
  {10.1146/annurev-astro-081309-130811}, \href
  {http://adsabs.harvard.edu/abs/2011ARA%26A..49..471M} {49, 471}

\bibitem[\protect\citeauthoryear{{Murray} \& {Dermott}}{{Murray} \&
  {Dermott}}{1999}]{MurrayDermott1999}
{Murray} C.~D.,  {Dermott} S.~F.,  1999, {Solar system dynamics}

\bibitem[\protect\citeauthoryear{{Musso Barcucci} et~al.,}{{Musso Barcucci}
  et~al.}{2019}]{MussoBarcucci2019}
{Musso Barcucci} A.,  et~al., 2019, \mn@doi [\aap]
  {10.1051/0004-6361/201935146}, \href
  {https://ui.adsabs.harvard.edu/abs/2019A&A...627A..77M} {627, A77}

\bibitem[\protect\citeauthoryear{{Mustill} \& {Wyatt}}{{Mustill} \&
  {Wyatt}}{2009}]{Mustill2009}
{Mustill} A.~J.,  {Wyatt} M.~C.,  2009, \mn@doi [\mnras]
  {10.1111/j.1365-2966.2009.15360.x}, \href
  {http://adsabs.harvard.edu/abs/2009MNRAS.399.1403M} {399, 1403}

\bibitem[\protect\citeauthoryear{{Pan}, {Nesvold}  \& {Kuchner}}{{Pan}
  et~al.}{2016}]{Pan2016}
{Pan} M.,  {Nesvold} E.~R.,   {Kuchner} M.~J.,  2016, \mn@doi [\apj]
  {10.3847/0004-637X/832/1/81}, \href
  {http://adsabs.harvard.edu/abs/2016ApJ...832...81P} {832, 81}

\bibitem[\protect\citeauthoryear{{Pearce} \& {Wyatt}}{{Pearce} \&
  {Wyatt}}{2014}]{Pearce2014}
{Pearce} T.~D.,  {Wyatt} M.~C.,  2014, \mn@doi [\mnras]
  {10.1093/mnras/stu1302}, \href
  {http://adsabs.harvard.edu/abs/2014MNRAS.443.2541P} {443, 2541}

\bibitem[\protect\citeauthoryear{{Pearce} \& {Wyatt}}{{Pearce} \&
  {Wyatt}}{2015}]{Pearce2015doublering}
{Pearce} T.~D.,  {Wyatt} M.~C.,  2015, \mn@doi [\mnras]
  {10.1093/mnras/stv1847}, \href
  {http://adsabs.harvard.edu/abs/2015MNRAS.453.3329P} {453, 3329}

\bibitem[\protect\citeauthoryear{{P{\'e}rez}, {Casassus}, {Baruteau}, {Dong},
  {Hales}  \& {Cieza}}{{P{\'e}rez} et~al.}{2019a}]{Perez2019}
{P{\'e}rez} S.,  {Casassus} S.,  {Baruteau} C.,  {Dong} R.,  {Hales} A.,
  {Cieza} L.,  2019a, \mn@doi [\aj] {10.3847/1538-3881/ab1f88}, \href
  {https://ui.adsabs.harvard.edu/abs/2019AJ....158...15P} {158, 15}

\bibitem[\protect\citeauthoryear{{P{\'e}rez}, {Marino}, {Casassus}, {Baruteau},
  {Zurlo}, {Flores}  \& {Chauvin}}{{P{\'e}rez}
  et~al.}{2019b}]{Perez2019protolunar}
{P{\'e}rez} S.,  {Marino} S.,  {Casassus} S.,  {Baruteau} C.,  {Zurlo} A.,
  {Flores} C.,   {Chauvin} G.,  2019b, \mn@doi [\mnras]
  {10.1093/mnras/stz1775}, \href
  {https://ui.adsabs.harvard.edu/abs/2019MNRAS.488.1005P} {488, 1005}

\bibitem[\protect\citeauthoryear{{Perrot} et~al.,}{{Perrot}
  et~al.}{2016}]{Perrot2016}
{Perrot} C.,  et~al., 2016, \mn@doi [\aap] {10.1051/0004-6361/201628396}, \href
  {https://ui.adsabs.harvard.edu/abs/2016A&A...590L...7P} {590, L7}

\bibitem[\protect\citeauthoryear{{Pinilla}, {Flock}, {Ovelar}  \&
  {Birnstiel}}{{Pinilla} et~al.}{2016}]{Pinilla2016}
{Pinilla} P.,  {Flock} M.,  {Ovelar} M. d.~J.,   {Birnstiel} T.,  2016, \mn@doi
  [\aap] {10.1051/0004-6361/201628441}, \href
  {https://ui.adsabs.harvard.edu/abs/2016A&A...596A..81P} {596, A81}

\bibitem[\protect\citeauthoryear{{Quillen}}{{Quillen}}{2006}]{Quillen2006}
{Quillen} A.~C.,  2006, \mn@doi [\mnras] {10.1111/j.1745-3933.2006.00216.x},
  \href {http://adsabs.harvard.edu/abs/2006MNRAS.372L..14Q} {372, L14}

\bibitem[\protect\citeauthoryear{{Rameau} et~al.,}{{Rameau}
  et~al.}{2016}]{Rameau2016}
{Rameau} J.,  et~al., 2016, \mn@doi [\apjl] {10.3847/2041-8205/822/2/L29},
  \href {http://adsabs.harvard.edu/abs/2016ApJ...822L..29R} {822, L29}

\bibitem[\protect\citeauthoryear{{Read}, {Wyatt}, {Marino}  \&
  {Kennedy}}{{Read} et~al.}{2018}]{Read2018}
{Read} M.~J.,  {Wyatt} M.~C.,  {Marino} S.,   {Kennedy} G.~M.,  2018, \mn@doi
  [\mnras] {10.1093/mnras/sty141}, \href
  {http://adsabs.harvard.edu/abs/2018MNRAS.475.4953R} {475, 4953}

\bibitem[\protect\citeauthoryear{{Reg{\'a}ly}, {Dencs}, {Mo{\'o}r}  \&
  {Kov{\'a}cs}}{{Reg{\'a}ly} et~al.}{2018}]{Regaly2018}
{Reg{\'a}ly} Z.,  {Dencs} Z.,  {Mo{\'o}r} A.,   {Kov{\'a}cs} T.,  2018, \mn@doi
  [\mnras] {10.1093/mnras/stx2604}, \href
  {http://adsabs.harvard.edu/abs/2018MNRAS.473.3547R} {473, 3547}

\bibitem[\protect\citeauthoryear{{Rein} \& {Liu}}{{Rein} \&
  {Liu}}{2012}]{rebound}
{Rein} H.,  {Liu} S.-F.,  2012, \mn@doi [\aap] {10.1051/0004-6361/201118085},
  \href {http://adsabs.harvard.edu/abs/2012A%26A...537A.128R} {537, A128}

\bibitem[\protect\citeauthoryear{{Rein} et~al.,}{{Rein}
  et~al.}{2019}]{mercurius}
{Rein} H.,  et~al., 2019, \mn@doi [\mnras] {10.1093/mnras/stz769}, \href
  {https://ui.adsabs.harvard.edu/abs/2019MNRAS.485.5490R} {485, 5490}

\bibitem[\protect\citeauthoryear{{Ricci}, {Carpenter}, {Fu}, {Hughes}, {Corder}
   \& {Isella}}{{Ricci} et~al.}{2015}]{Ricci2015}
{Ricci} L.,  {Carpenter} J.~M.,  {Fu} B.,  {Hughes} A.~M.,  {Corder} S.,
  {Isella} A.,  2015, \mn@doi [\apj] {10.1088/0004-637X/798/2/124}, \href
  {http://adsabs.harvard.edu/abs/2015ApJ...798..124R} {798, 124}

\bibitem[\protect\citeauthoryear{{Richert}, {Lyra}  \& {Kuchner}}{{Richert}
  et~al.}{2018}]{Richert2017}
{Richert} A.~J.~W.,  {Lyra} W.,   {Kuchner} M.~J.,  2018, \mn@doi [\apj]
  {10.3847/1538-4357/aaadaa}, \href
  {http://adsabs.harvard.edu/abs/2018ApJ...856...41R} {856, 41}

\bibitem[\protect\citeauthoryear{{Saito} \& {Sirono}}{{Saito} \&
  {Sirono}}{2011}]{Saito2011}
{Saito} E.,  {Sirono} S.-i.,  2011, \mn@doi [\apj]
  {10.1088/0004-637X/728/1/20}, \href
  {https://ui.adsabs.harvard.edu/abs/2011ApJ...728...20S} {728, 20}

\bibitem[\protect\citeauthoryear{{Schneider} et~al.,}{{Schneider}
  et~al.}{2014}]{Schneider2014}
{Schneider} G.,  et~al., 2014, \mn@doi [\aj] {10.1088/0004-6256/148/4/59},
  \href {http://adsabs.harvard.edu/abs/2014AJ....148...59S} {148, 59}

\bibitem[\protect\citeauthoryear{{Shannon}, {Mustill}  \& {Wyatt}}{{Shannon}
  et~al.}{2015}]{Shannon2015}
{Shannon} A.,  {Mustill} A.~J.,   {Wyatt} M.,  2015, \mn@doi [\mnras]
  {10.1093/mnras/stv045}, \href
  {https://ui.adsabs.harvard.edu/abs/2015MNRAS.448..684S} {448, 684}

\bibitem[\protect\citeauthoryear{{Simpson} et~al.,}{{Simpson}
  et~al.}{2015}]{Simpson2015size}
{Simpson} J.~M.,  et~al., 2015, \mn@doi [\apj] {10.1088/0004-637X/799/1/81},
  \href {http://adsabs.harvard.edu/abs/2015ApJ...799...81S} {799, 81}

\bibitem[\protect\citeauthoryear{{Smith} \& {Lissauer}}{{Smith} \&
  {Lissauer}}{2009}]{Smith2009stability}
{Smith} A.~W.,  {Lissauer} J.~J.,  2009, \mn@doi [\icarus]
  {10.1016/j.icarus.2008.12.027}, \href
  {http://adsabs.harvard.edu/abs/2009Icar..201..381S} {201, 381}

\bibitem[\protect\citeauthoryear{{Stolker} et~al.,}{{Stolker}
  et~al.}{2019}]{Stolker2020}
{Stolker} T.,  et~al., 2019, arXiv e-prints, \href
  {https://ui.adsabs.harvard.edu/abs/2019arXiv191213316S} {p. arXiv:1912.13316}

\bibitem[\protect\citeauthoryear{{Su} et~al.,}{{Su} et~al.}{2017}]{Su2017}
{Su} K.~Y.~L.,  et~al., 2017, \mn@doi [\aj] {10.3847/1538-3881/aa906b}, \href
  {http://adsabs.harvard.edu/abs/2017AJ....154..225S} {154, 225}

\bibitem[\protect\citeauthoryear{{Takahashi} \& {Inutsuka}}{{Takahashi} \&
  {Inutsuka}}{2014}]{Takahashi2014}
{Takahashi} S.~Z.,  {Inutsuka} S.-i.,  2014, \mn@doi [\apj]
  {10.1088/0004-637X/794/1/55}, \href
  {http://adsabs.harvard.edu/abs/2014ApJ...794...55T} {794, 55}

\bibitem[\protect\citeauthoryear{{Visser}, {van Dishoeck}  \& {Black}}{{Visser}
  et~al.}{2009}]{Visser2009}
{Visser} R.,  {van Dishoeck} E.~F.,   {Black} J.~H.,  2009, \mn@doi [\aap]
  {10.1051/0004-6361/200912129}, \href
  {http://adsabs.harvard.edu/abs/2009A%26A...503..323V} {503, 323}

\bibitem[\protect\citeauthoryear{{Vorobyov}}{{Vorobyov}}{2013}]{Vorobyov2013}
{Vorobyov} E.~I.,  2013, \mn@doi [\aap] {10.1051/0004-6361/201220601}, \href
  {https://ui.adsabs.harvard.edu/abs/2013A&A...552A.129V} {552, A129}

\bibitem[\protect\citeauthoryear{{Vousden}, {Farr}  \& {Mandel}}{{Vousden}
  et~al.}{2016}]{Vousden2016}
{Vousden} W.~D.,  {Farr} W.~M.,   {Mandel} I.,  2016, \mn@doi [\mnras]
  {10.1093/mnras/stv2422}, \href
  {https://ui.adsabs.harvard.edu/abs/2016MNRAS.455.1919V} {455, 1919}

\bibitem[\protect\citeauthoryear{{Wilner}, {MacGregor}, {Andrews}, {Hughes},
  {Matthews}  \& {Su}}{{Wilner} et~al.}{2018}]{Wilner2018}
{Wilner} D.~J.,  {MacGregor} M.~A.,  {Andrews} S.~M.,  {Hughes} A.~M.,
  {Matthews} B.,   {Su} K.,  2018, \mn@doi [\apj] {10.3847/1538-4357/aaacd7},
  \href {http://adsabs.harvard.edu/abs/2018ApJ...855...56W} {855, 56}

\bibitem[\protect\citeauthoryear{{Wisdom}}{{Wisdom}}{1980}]{Wisdom1980}
{Wisdom} J.,  1980, \mn@doi [\aj] {10.1086/112778}, \href
  {http://adsabs.harvard.edu/abs/1980AJ.....85.1122W} {85, 1122}

\bibitem[\protect\citeauthoryear{{Wyatt}}{{Wyatt}}{2008}]{Wyatt2008}
{Wyatt} M.~C.,  2008, \mn@doi [\araa] {10.1146/annurev.astro.45.051806.110525},
  \href {http://adsabs.harvard.edu/abs/2008ARA%26A..46..339W} {46, 339}

\bibitem[\protect\citeauthoryear{{Wyatt} \& {Dent}}{{Wyatt} \&
  {Dent}}{2002}]{Wyatt2002}
{Wyatt} M.~C.,  {Dent} W.~R.~F.,  2002, \mn@doi [\mnras]
  {10.1046/j.1365-8711.2002.05533.x}, \href
  {http://adsabs.harvard.edu/abs/2002MNRAS.334..589W} {334, 589}

\bibitem[\protect\citeauthoryear{{Wyatt}, {Greaves}, {Dent}  \&
  {Coulson}}{{Wyatt} et~al.}{2005}]{Wyatt2005}
{Wyatt} M.~C.,  {Greaves} J.~S.,  {Dent} W.~R.~F.,   {Coulson} I.~M.,  2005,
  \mn@doi [\apj] {10.1086/426929}, \href
  {http://adsabs.harvard.edu/abs/2005ApJ...620..492W} {620, 492}

\bibitem[\protect\citeauthoryear{{Wyatt}, {Smith}, {Greaves}, {Beichman},
  {Bryden}  \& {Lisse}}{{Wyatt} et~al.}{2007}]{Wyatt2007hotdust}
{Wyatt} M.~C.,  {Smith} R.,  {Greaves} J.~S.,  {Beichman} C.~A.,  {Bryden} G.,
   {Lisse} C.~M.,  2007, \mn@doi [\apj] {10.1086/510999}, \href
  {http://adsabs.harvard.edu/abs/2007ApJ...658..569W} {658, 569}

\bibitem[\protect\citeauthoryear{{Wyatt}, {Booth}, {Payne}  \&
  {Churcher}}{{Wyatt} et~al.}{2010}]{Wyatt2010}
{Wyatt} M.~C.,  {Booth} M.,  {Payne} M.~J.,   {Churcher} L.~J.,  2010, \mn@doi
  [\mnras] {10.1111/j.1365-2966.2009.15930.x}, \href
  {http://adsabs.harvard.edu/abs/2010MNRAS.402..657W} {402, 657}

\bibitem[\protect\citeauthoryear{{Wyatt} et~al.,}{{Wyatt}
  et~al.}{2012}]{Wyatt2012}
{Wyatt} M.~C.,  et~al., 2012, \mn@doi [\mnras]
  {10.1111/j.1365-2966.2012.21298.x}, \href
  {http://adsabs.harvard.edu/abs/2012MNRAS.424.1206W} {424, 1206}

\bibitem[\protect\citeauthoryear{{Wyatt}, {Kral}  \& {Sinclair}}{{Wyatt}
  et~al.}{2020}]{Wyatt2020}
{Wyatt} M.~C.,  {Kral} Q.,   {Sinclair} C.~A.,  2020, \mn@doi [\mnras]
  {10.1093/mnras/stz3052}, \href
  {https://ui.adsabs.harvard.edu/abs/2020MNRAS.491..782W} {491, 782}

\bibitem[\protect\citeauthoryear{{Yelverton} \& {Kennedy}}{{Yelverton} \&
  {Kennedy}}{2018}]{Yelverton2018}
{Yelverton} B.,  {Kennedy} G.~M.,  2018, \mn@doi [\mnras]
  {10.1093/mnras/sty1678}, \href
  {http://cdsads.u-strasbg.fr/abs/2018MNRAS.479.2673Y} {479, 2673}

\bibitem[\protect\citeauthoryear{{Yelverton}, {Kennedy}, {Su}  \&
  {Wyatt}}{{Yelverton} et~al.}{2019}]{Yelverton2019}
{Yelverton} B.,  {Kennedy} G.~M.,  {Su} K. Y.~L.,   {Wyatt} M.~C.,  2019,
  \mn@doi [\mnras] {10.1093/mnras/stz1927}, \href
  {https://ui.adsabs.harvard.edu/abs/2019MNRAS.488.3588Y} {488, 3588}

\bibitem[\protect\citeauthoryear{{Yelverton}, {Kennedy}  \& {Su}}{{Yelverton}
  et~al.}{2020}]{Yelverton2020}
{Yelverton} B.,  {Kennedy} G.~M.,   {Su} K. Y.~L.,  2020, arXiv e-prints, \href
  {https://ui.adsabs.harvard.edu/abs/2020arXiv200503573Y} {p. arXiv:2005.03573}

\bibitem[\protect\citeauthoryear{{Zapata}, {Ho}  \& {Rodr{\'\i}guez}}{{Zapata}
  et~al.}{2018}]{Zapata2018}
{Zapata} L.~A.,  {Ho} P. T.~P.,   {Rodr{\'\i}guez} L.~F.,  2018, \mn@doi
  [\mnras] {10.1093/mnras/sty420}, \href
  {https://ui.adsabs.harvard.edu/abs/2018MNRAS.476.5382Z} {476, 5382}

\bibitem[\protect\citeauthoryear{{Zuckerman} \& {Song}}{{Zuckerman} \&
  {Song}}{2012}]{Zuckerman2012}
{Zuckerman} B.,  {Song} I.,  2012, \mn@doi [\apj] {10.1088/0004-637X/758/2/77},
  \href {http://adsabs.harvard.edu/abs/2012ApJ...758...77Z} {758, 77}

\bibitem[\protect\citeauthoryear{{Zuckerman}, {Forveille}  \&
  {Kastner}}{{Zuckerman} et~al.}{1995}]{Zuckerman1995}
{Zuckerman} B.,  {Forveille} T.,   {Kastner} J.~H.,  1995, \mn@doi [\nat]
  {10.1038/373494a0}, \href {http://adsabs.harvard.edu/abs/1995Natur.373..494Z}
  {373, 494}

\bibitem[\protect\citeauthoryear{{Zurlo} et~al.,}{{Zurlo}
  et~al.}{2016}]{Zurlo2016}
{Zurlo} A.,  et~al., 2016, \mn@doi [\aap] {10.1051/0004-6361/201526835}, \href
  {https://ui.adsabs.harvard.edu/abs/2016A&A...587A..57Z} {587, A57}

\bibitem[\protect\citeauthoryear{{van Leeuwen}}{{van
  Leeuwen}}{2007}]{vanLeeuwen2007}
{van Leeuwen} F.,  2007, \mn@doi [\aap] {10.1051/0004-6361:20078357}, \href
  {http://cdsads.u-strasbg.fr/abs/2007A%26A...474..653V} {474, 653}

\bibitem[\protect\citeauthoryear{{van Lieshout}, {Dominik}, {Kama}  \&
  {Min}}{{van Lieshout} et~al.}{2014}]{vanLieshout2014}
{van Lieshout} R.,  {Dominik} C.,  {Kama} M.,   {Min} M.,  2014, \mn@doi [\aap]
  {10.1051/0004-6361/201322090}, \href
  {http://adsabs.harvard.edu/abs/2014A%26A...571A..51V} {571, A51}

\makeatother
\end{thebibliography}




\appendix

\section{Continuum imaging}
\label{appendix}
\begin{figure*}
  \centering \includegraphics[trim=0.0cm 0.0cm 0.0cm 0.0cm, clip=true,
    width=0.45\textwidth]{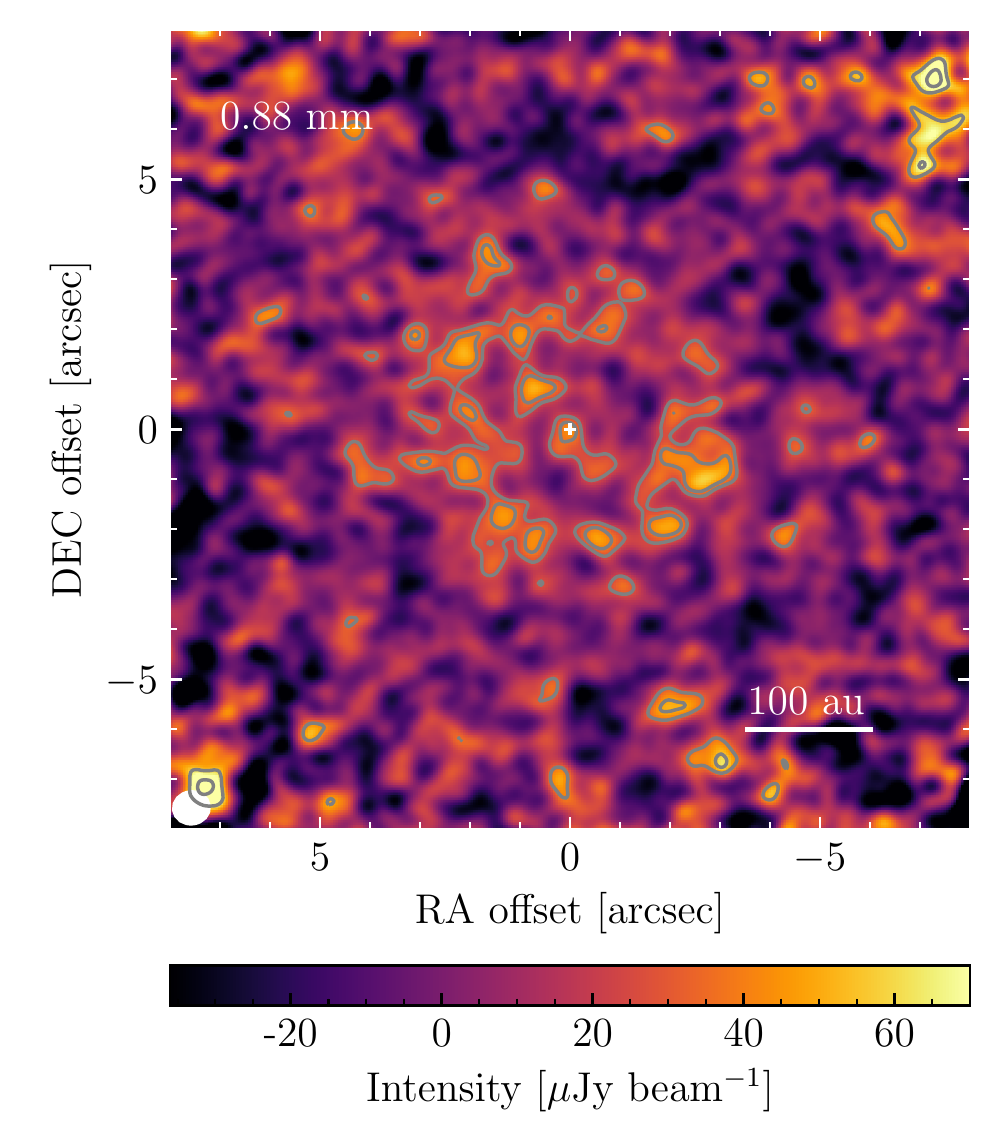}
  \includegraphics[trim=0.0cm 0.0cm 0.0cm 0.0cm, clip=true,
    width=0.45\textwidth]{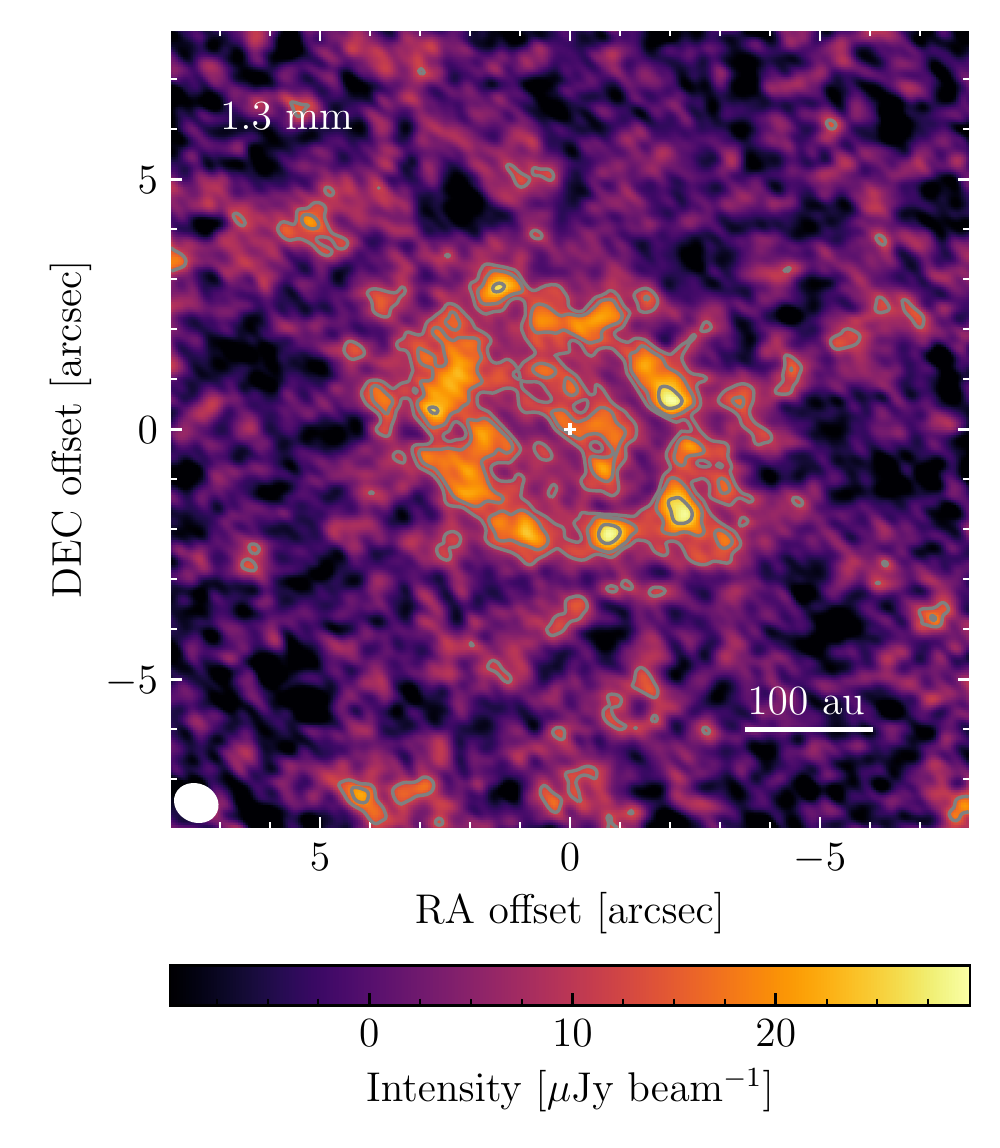}
 \caption{Continuum Clean images at 0.88~mm (12m+ACA, left panel) and
   1.3~mm (right panel) of HD~206893 obtained using Briggs weighting
   and a robust parameter of 2. Additionally, we applied a uv tapering
   of $0\farcs4$ to the band 7 data. The images are also corrected by
   the primary beam, hence the noise increases towards the edges. The
   contours represent 2, 3 and 5 times the image rms (12 and
   4.9~\uJybeam\ at the center of the band 7 and 6 images,
   respectively). The stellar position is marked with a white cross
   near the center of the image (based on Gaia DR2) and the beams are
   represented by white ellipses in the bottom left corners
   ($0\farcs57\times0\farcs50$ and $0\farcs70\times0\farcs57$,
   respectively). }
 \label{fig:continuum_hr}
\end{figure*}

In Figure \ref{fig:continuum_hr} we present the continuum clean images
used to compute the radial profiles in Figure~\ref{fig:radial}.

\bsp	
\label{lastpage}
\end{document}